\documentclass[preprint,showpacs]{revtex4}
\usepackage{epsfig}

\begin{document}

\title{Systematic study of multi-quark states I. $qq-qq-\bar{q}$ configuration}
\author{Hongxia Huang, Chengrong Deng, Jialun Ping}

\address{Physics Department, Nanjing Normal University, Nanjing, 210097
P.R. China}

\author{Fan Wang}

\address{Center for Theoretical Physics, Nanjing University, Nanjing 210093,
P.R. China}

\author{T. Goldman}

\address{Theoretical Division, LANL, Los Alamos, NM 87545, USA}

\begin{abstract}
Group theoretic method for the systematic study of multi-quark
states is developed. The calculation of matrix elements of many-body
Hamiltonian is simplified by transforming the physical bases (quark
cluster bases) to symmetry bases (group chain classified bases),
where the fractional parentage expansion method can be used.
Five-quark system is taken as an example in this study. The
Jaffe-Wilczek $qq-qq-\bar{q}$ configuration is chosen as one of
examples to construct the physical bases and the transformation
coefficients between physical bases and symmetry ones are shown to
be related to the ${SU}_{mn}\supset{SU}_m\times{SU}_n$ isoscalar
factors and a complete transformation coefficients table is given.
The needed isoscalar factors and fractional parentage coefficients
had been calculated with our new group representation theory and
published before. Three quark models: the naive Glashow-Isgur,
Salamanca and quark delocalization color screening, are used to show
the general applicability of the new multi-quark calculation method
and general results of constituent quark models for five-quark
states are given.
\end{abstract}


\maketitle

\section{Introduction}

Hadron (baryons and mesons) spectroscopy opens the gate for the
development of the fundamental theory of the strong interaction:
quantum chromodynamics (QCD). However the non-perturbative
complication of low energy QCD makes it impossible to calculate the
hadron structure analytically from QCD directly. The unique color
structure of the known hadrons makes the construction of quark
models very efficient (Fig.1). A variety of quark models employing
two-body interactions give a good description of hadron properties.
(For baryon, the three-body interaction can be well approximated by
two-body interaction.) However the unique color structure also
limits our understanding of the properties of other color structures
available in QCD. In order to understand low energy QCD, to study
system with more quarks is indispensable. Hadron-hadron scattering
provides a window on the nature of other color structures, but it is
not enough because the colorless meson exchange model and chiral
perturbation theory both describe the low energy hadron-hadron
scattering data well. QCD does not rule out the existence of
glueballs, quark-gluon hybrids, multi-quark states, {\em etc.} based
on the present understanding. Multi-quark systems are important
samples for providing information on low energy QCD interaction,
especially for complex color structures (Fig.2).
\begin{center}

\epsfxsize=3.5in \epsfbox{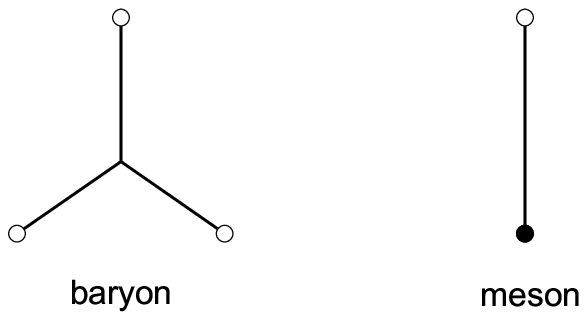}

Fig.1 The color structure of baryons and mesons.
\end{center}

Since 1977, interest in multi-quark systems has continued although
there have been rising and falling. The search for the H-particle
(6-quark system), which was predicted in 1977 by R. Jaffe
\cite{Jaffe} with MIT bag model, has not succeeded for almost 30
years. In 1993, the dibaryon state $d'$ appeared unexpectedly
\cite{dprime}, observed in double charge exchange reactions but
disappeared nine years later. Other six quark states, :$d^*$
\cite{dibaryon}, $N\Omega$ \cite{nomega} and $\Omega\Omega$
\cite{diomega} {\em etc.}, had been proposed but none of them has
been established experimentally. In 2003, the pentaquark state,
$\Theta^+$, aroused a new enthusiasm in multi-quark systems. More
than ten groups claimed that they observed the $\Theta^+$ signal,
but almost the same number of groups did not \cite{penta1,penta2}.
Its appearance had raised great trouble for theory of hadron
spectroscopy, almost none of the models constrained by hadron
properties and hadron-hadron scattering can account for the
$\Theta^+$ \cite{penta3,pwg}. In studying pentaquark $\Theta^+$,
various color and spatial structures of pentaquark had been
proposed: color singlet hadron molecules [K($q\bar{q}$)N($q^3$)],
color anti-triplet diquarks [$(qq)(qq)\bar{q}$] \cite{JW},
diquark-triquark [$qq-qq\bar{q}$] \cite{KL}, quark methane
[$q^4\bar{q}$] \cite{SZ}, {\em etc}. However recent high statistic
experiments did not confirm the $\Theta^+$ signal \cite{JLab}.
Today, about three years later, pentaquark $\Theta^+$ seems to be
about to disappear also.

After the pentaquark $\Theta^+$ (a member of anti-decuplet), other
states in 27-plet and 35-plet \cite{27plet,35plet,Bijker}, had been
proposed. In order to understand the nucleon spin structure within
constituent quark model the five-quark component is necessary
\cite{qcw}. Recently to explain the strange magnetic moment of
proton, five-quark component was introduced in the nucleon
\cite{BSZou}.

Tetra-quark states are re-interested both experimentally and
theoretically because of new discoveries since 2003
\cite{tetraquark}.

\begin{center}
\epsfxsize=4.5in \epsfbox{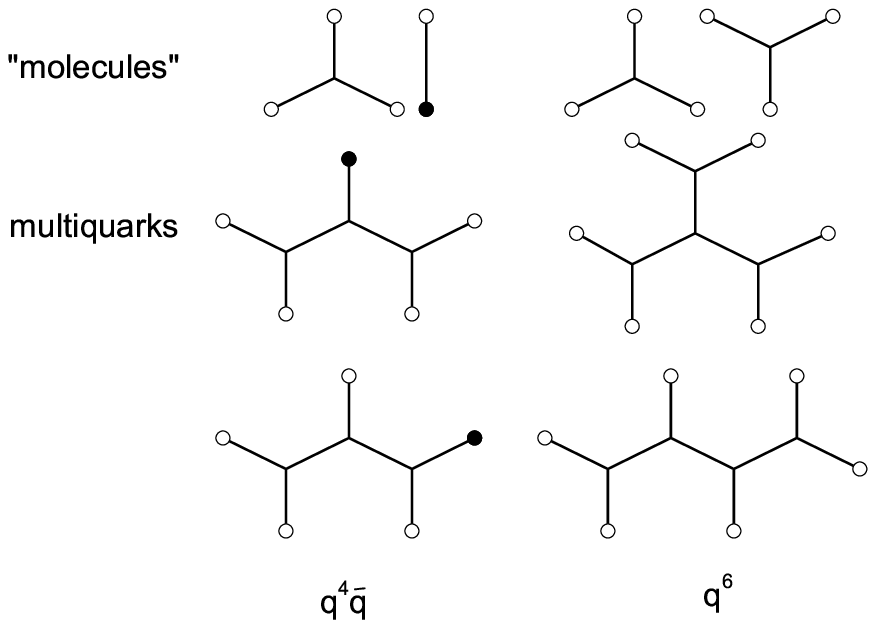}

Fig.2 The color structure of multi-quark states.
\end{center}

Fig.2 shows possible color structures for 5- and 6-quark systems
based on the lattice QCD calculation \cite{lat} and general color
confinement idea. Clearly they have more possible color structures
than 2- and 3-quark hadrons. Due to color confinement, only color
singlet combinations of quarks can be separated. Colorful clusters
will be confined in a genuine multi-quark system. It will transit to
color singlet sub-systems through color flux rearrangement first
then decay. This will induces resonance similar to compound nucleus
formation but due to color confinement and could be called color
confinement resonance. To include these intermediate hidden color
configurations in the quark model, a multi-channel coupling
calculation with multi-body interactions is required. This is quite
involved \cite{vji} and finding a method to make the calculation
tractable is an important element in the study of multi-quark
systems with constituent quark models. Moreover up to now we almost
don't have any idea about the transition interaction between
different color configurations. To develop model to include effects
of these hidden color configurations is expected.

The group theory classification of 4, 5 and 6 quark states had been
published \cite{ES,Bijker,qdcsm3}. The fractional parentage (fp)
expansion technique has been proven as a powerful method for
few-body problem. To full play the powerful group theoretic method
for quark model calculation one needs not only the fractional
parentage expansion coefficients of these multi quark states but
also a relation between various quark model states (hereafter called
physical bases) and the group theoretical classification states
(hereafter called symmetry bases). Such a method has been developed
and successfully applied in systematic search of dibaryon
\cite{qdcsm3,qdcsm4,gpw}, where the physical bases were transformed
to symmetry bases first, then the 6-body matrix elements calculation
of Hamiltonian (with two body interaction) on the symmetry bases was
done which can be reduced to four-body overlap and two-body matrix
elements calculation by means of fractional parentage expansion. At
last the matrix elements on the symmetry bases were transformed back
to physical bases.

The main content of this paper is to provide the transformation
coefficients between physical bases and symmetry bases of five-quark
systems to facilitate the calculation of many-body Hamiltonian
matrix elements. The physical bases discussed in this paper is the
Jaffe-Wilczek diquark model ones. Another useful physical bases is
the meson-baryon bases which will be given in a company paper. To
illustrate the application of this group theory method for
five-quark systems three quark models were employed for pentaquark
calculation. They are the naive quark model, i.e., the Glashow-Isgur
model \cite{Isgur}; the Salamanca chiral quark model \cite{chiral}
and the quark delocalization color screening model (QDCSM) developed
by our group \cite{qdcsm1}. The calculation of the related
fractional parentage coefficients will be mentioned but the needed
results have been published elsewhere \cite{chen,ISFBook}.

In section II, the physical bases and symmetry bases are introduced
and the transformation between them is derived. The fractional
parentage technique applied to calculate the matrix elements on the
symmetry bases is also explained in this section. Section III
explains three quark models we used. The results of the systematic
calculation of pentaquark in the $u,d,s$ 3-flavor world are given in
section IV. The last section gives the summary.

\section{Physical bases and symmetry bases}

The physical bases are constructed as follows, first the wave
function of each quark cluster was constructed based on the group
chain classification
\[{SU}_{36}\supset{SU}^x_2\times\left\{{SU}_{18}\supset
{SU}^c_3\times\left[{SU}_6\supset\left({SU}^f_3\supset
{SU}^{\tau}_2\times{U}^Y_1\right)\times{SU}^{\sigma}_2\right]\right\},\]
then the quark cluster wave functions of the system was coupled to
definite color, spin and isospin quantum numbers by Clebsch-Gordan
(CG) coefficients of color $SU^c_3$, spin $SU^{\sigma}_2$ and
isospin $SU^{\tau}_2$ group, and finally anti-symmetrized.

\begin{center}
\epsfxsize=2.5in \epsfbox{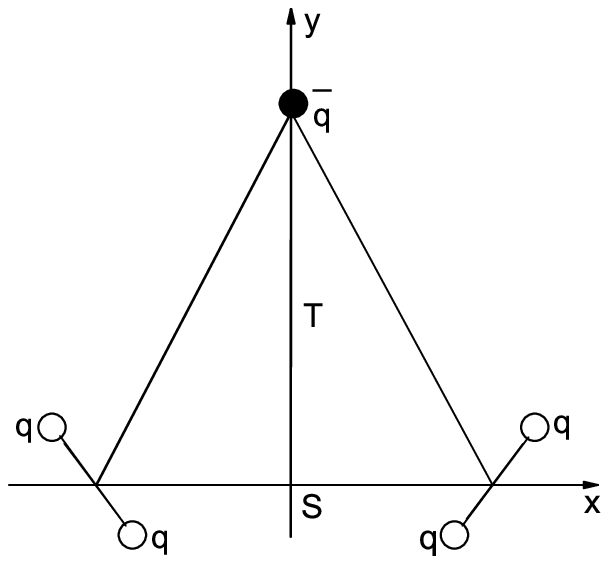}

Fig.3 The Jaffe-Wilczek configuration of pentaquark.
\end{center}

For Jaffe-Wilczek di-quark configuration, the five quarks are
separated into three clusters and form an isosceles triangle with
the two strongly correlated pairs of quarks sitting at the bottom
corners with separation $S$ and the anti-quark at top with the
height $T$ (see Fig.3). The di-quark in the $u,d,s$ three-flavor
world is described by
\begin{equation}
\psi_2(q^2)=\left| \begin{array}{c} [\nu_2]W_{\nu_2} \\ \left[ c_2
\right] W_{c_2} [ \mu_2][f_2]Y_2I_2M_{I_2}J_2M_{J_2}
\end{array} \right\rangle
\end{equation}
which is the basis vector belonging to the irreducible
representations ($n=2$)
\begin{eqnarray}
& & ~[1^n]~~~~~~[\nu]~~~~~~~ [{\tilde
\nu}]~~~~~~~[c]~~~~~~~[\mu]~~~~~~~
[f] ~~~~~~~I~~~~~~Y~~~~~~~~J~~ \nonumber \\
& & \mbox{SU}_{36}\supset \mbox{SU}^x_2\times \left\{
\mbox{SU}_{18}\supset \mbox{SU}^c_3\times \left[ \mbox{SU}_6\supset
\left( \mbox{SU}^f_3\supset \mbox{SU}^{\tau}_2\times \mbox{U}^Y_1
\right) \times \mbox{SU}^{\sigma}_2 \right] \right\}
\label{group-chain}
\end{eqnarray}
where $[\nu ]$, W, {\em etc.} are the Young diagrams, Weyl tableaux,
{\em etc.} describing the permutation and SU$_n$ symmetries. In our
calculation, the ground state diquarks are assumed to be in the
totally symmetric orbital state $[\nu_2]=[2]$. The cluster basis for
4-quark can be defined as
\begin{equation}
\Psi_{\alpha_4 k_4}(q^4) = {\cal A} \left[ \psi_2(q_1q_2)
\psi_2(q_3q_4) \right]^{[c_4 ]I_4J_4}_{W_{c_4}M_{I_4}M_{J_4}},
\end{equation}
where ${\cal A}$ is a normalized antisymmetric operator for 4-quark,
$[~]$ means coupling in terms of the SU$_3^c$, SU$_2^\tau$,
SU$_2^\sigma$ CG coefficients so that it has color symmetry
$[c_4]W_{c_4}$, isospin $I_4M_{I_4}$, and spin $J_4M_{J_4}$.
$\alpha_4=(Y_4I_4J_4)$, $k_4$ represents the quantum numbers
$\nu_2,\nu'_2,c_2,\cdots,J'_2$. The cluster basis for 5-quark can be
obtained by coupling the antiquark basis to 4-quark basis in terms
of the SU$_3^c$, SU$_2^\tau$, SU$_2^\sigma$ CG coefficients,
\begin{equation}
\Psi_{\alpha k}(q^4\bar{q}) = \left[ \Psi_{\alpha_4 k_4}(q^4))
\psi_{[\bar{c}]\bar{I}\bar{J}}(\bar{q}_5) \right]^{[c]IJ}_{WM_IM_J}.
\label{phys_q5}
\end{equation}

The symmetry basis of 4-quark system is just the group chain
(Eq.(\ref{group-chain})) classification basis ($n=4$),
\begin{equation}
\Phi_{\alpha_4 K_4}(q^4) = \left| \begin{array}{c} [\nu_4]W_{\nu_4}
\\ \left[ c_4 \right] W_{c_4} [ \mu_4][f_4]Y_4I_4M_{I_4}J_4M_{J_4}
\end{array} \right\rangle,
\end{equation}
where $K_4$ stands for $\nu_4\mu_4f_4$. Similar to
Eq.(\ref{phys_q5}), the symmetry bases for 5-quark is
\begin{equation}
\Phi_{\alpha K}(q^4\bar{q}) = \left[ \Phi_{\alpha_4 K_4}(q^4))
\psi_{[\bar{c}]\bar{I}\bar{J}}(\bar{q}_5) \right]^{[c]IJ}_{WM_IM_J}.
\label{symm_q5}
\end{equation}
The cluster bases and symmetry bases for 5-quark can be transformed
to each other \cite{Harvey,chen1,qdcsm3,qdcsm4}.
\begin{eqnarray}
\Psi_{\alpha k}(q^4\bar{q}) & = & \sum_{K} C_{k K} \Phi_{\alpha
K}(q^4\bar{q}) \nonumber \\
& = & \sum_{\tilde{\nu}_4\mu_4f_4}
C^{[\tilde{\nu}_4][c_4][\mu_4]}_{[\tilde{\nu}_2][c_2][\mu_2],
[\tilde{\nu}'_2][c'_2][\mu'_2]}
C^{[\mu_4][f_4][J_4]}_{[\mu_2][f_2][J_2],[\mu'_2][f'_2][J'_2]}
C^{[f_4]Y_4I_4}_{[f_2]Y_2I_2,[f'_2]Y'_2I'_2} \Phi_{\alpha
K}(q^4\bar{q}), \label{trans}
\end{eqnarray}
$C'$s are the isoscalar factors of
${SU}_{mn}\supset{SU}_m\times{SU}_n$, which can be obtained from the
book \cite{ISFBook}. All the transformation coefficients are listed
in the appendix.

A physical 5-quark state with quantum number $\alpha =(YIJ)$ is
expressed as a channel coupling wave function
\begin{equation}
\Psi_{\alpha}(q^4\bar{q}) = \sum_k C_k \Psi_{\alpha k}(q^4\bar{q}).
\end{equation}
The channel coupling coefficient $C_k$ is determined by the
diagonalization of the 5-quark Hamiltonian as usual. The calculation
of Hamiltonian matrix elements in the cluster bases is tedious and
it can be replaced by the matrix elements in the symmetry bases by
the transformation Eq.(\ref{trans}),
\begin{equation}
\langle \Psi_{\alpha k} | H |
 \Psi _{\alpha k^{\prime}} \rangle =\sum_{K,K^{\prime}} C_{k K}
C_{k' K^{\prime}} \left\langle \Phi_{\alpha K}(q^4\bar{q}) \right| H
\left| \Phi_{\alpha K'}(q^4\bar{q}) \right\rangle .
\end{equation}
In the symmetry bases, the matrix elements $\langle \Phi _{\alpha
K}\left| H\right|\Phi _{\alpha K^{\prime}} \rangle $ can be
calculated by the well known fp expansion method. Because there is
an antiquark, we have to use different fp expansion for $qq$
interaction and $q\bar{q}$ interaction. For $qq$ interaction,
$4\rightarrow 2+2$ is used.
\begin{eqnarray}
\langle \Phi_{\alpha K} \left| H_{34} \right| \Phi _{\alpha
K^{\prime}} \rangle & = &  \sum_{1,2}
C^{[1^4][\nu_4][\tilde{\nu}_4]}_{[1^2][\nu_1][\tilde{\nu}_1],[1^2][\nu_2][\tilde{\nu}_2]}
C^{[\tilde{\nu}_4][c_4][\mu_4]}_{[\tilde{\nu}_1][c_1][\mu_1],[\tilde{\nu}_2][c_2][\mu_2]}
C^{[\mu_4][f_4]J_4}_{[\mu_1][f_1]J_1,[\mu_2][f_2]J_2}
C^{[f_4]Y_4I_4}_{[f_1]Y_1I_1,[f_2]Y_2I_2}    \nonumber\\
& &
C^{[1^4][\nu_4][\tilde{\nu}_4]}_{[1^2][\nu'_1][\tilde{\nu}'_1],[1^2][\nu'_2][\tilde{\nu}'_2]}
C^{[\tilde{\nu}_4][c_4][\mu_4]}_{[\tilde{\nu}'_1][c'_1][\mu'_1],[\tilde{\nu}'_2][c'_2][\mu'_2]}
C^{[\mu_4][f_4]J_4}_{[\mu'_1][f'_1]J'_1,[\mu'_2][f'_2]J'_2}
C^{[f_4]Y_4I_4}_{[f'_1]Y'_1I'_1,[f'_2]Y'_2I'_2}     \nonumber\\
& & C^{[\nu_4]W_{x_4}}_{[\nu_1]W_{x_1},[\nu_2]W_{x_2}}
C^{[\nu_4]W_{x'_4}}_{[\nu'_1]W_{x'_1},[\nu'_2]W_{x'_2}} \langle
\alpha_1 K_1 | \alpha^{\prime}_1 K^{\prime}_1 \rangle \langle
\alpha_2 K_2 |H_{34} | \alpha^{\prime}_2 K^{\prime}_2 \rangle .
\end{eqnarray}
Here $C'$s are the ${SU}_{mn}\supset{SU}_m\times{SU}_n$ isoscalar
factors, $\langle \alpha_1 K_1 | \alpha^{\prime}_1 K^{\prime}_1
\rangle $ is the two quark overlap and $\langle \alpha_2 K_2 |H_{34}
| \alpha^{\prime}_2 K^{\prime}_2 \rangle $ is the two body matrix
element and $H_{34}$ represents two-body operator for the second
pair. The interacting pair number is $C^4_2=6$. For $q\bar{q}$
interaction, $4\rightarrow 3+1$ is used,
\begin{eqnarray}
\langle \Phi_{\alpha K} \left| H_{15} \right| \Phi _{\alpha
K^{\prime}} \rangle & = & \sum_{3,1}
C^{[1^4][\nu_4][\tilde{\nu}_4]}_{[1^3][\nu_3][\tilde{\nu}_3],[1][\nu_1][\tilde{\nu}_1]}
C^{[\tilde{\nu}_4][c_4][\mu_4]}_{[\tilde{\nu}_3][c_3][\mu_3],[\tilde{\nu}_1][c_1][\mu_1]}
C^{[\mu_4][f_4]J_4}_{[\mu_3][f_3]J_3,[\mu_1][f_1]J_1}
C^{[f_4]Y_4I_4}_{[f_3]Y_3I_3,[f_1]Y_1I_1}    \nonumber\\
& &
C^{[1^4][\nu_4][\tilde{\nu}_4]}_{[1^3][\nu'_3][\tilde{\nu}'_3],[1][\nu'_1][\tilde{\nu}'_1]}
C^{[\tilde{\nu}_4][c_4][\mu_4]}_{[\tilde{\nu}'_3][c'_3][\mu'_3],[\tilde{\nu}'_1][c'_1][\mu'_1]}
C^{[\mu_4][f_4]J_4}_{[\mu'_3][f'_3]J'_3,[\mu'_1][f'_1]J'_1}
C^{[f_4]Y_4I_4}_{[f'_3]Y'_3I'_3,[f'_1]Y'_1I'_1}     \nonumber\\
& & U(c_3c_1cc_{\bar{1}};c_4c_2)
U(I_3I_1II_{\bar{1}};I_4I_2)U(J_3J_1JJ_{\bar{1}};J_4J_2)  \nonumber\\
& & U(c'_3c'_1cc'_{\bar{1}};c_4c'_2)
U(I'_3I'_1II'_{\bar{1}};I_4I'_2)U(J'_3J'_1JJ'_{\bar{1}};J_4J'_2)  \nonumber\\
& & C^{[\nu_4]W_{x_4}}_{[\nu_3]W_{x_3},[\nu_1]W_{x_1}}
C^{[\nu_4]W_{x'_4}}_{[\nu'_3]W_{x'_3},[\nu'_1]W_{x'_1}} \langle
\alpha_3 K_3 | \alpha^{\prime}_3 K^{\prime}_3 \rangle \langle
\alpha_2 K_2 |H_{45} | \alpha^{\prime}_2 K^{\prime}_2 \rangle .
\end{eqnarray}
Here $U'$s are Racah coefficients, $\langle \alpha_3 K_3 |
\alpha^{\prime}_3 K^{\prime}_3 \rangle $ is the three quark overlap
and $\langle \alpha_2 K_2 |H_{45} | \alpha^{\prime}_2 K^{\prime}_2
\rangle $ is the two body matrix element and $H_{45}$ represents
quark-antiquark operator. The interacting pair number is $C^4_1=4$.
The calculation of every factor in above equations can be found in
Ref.\cite{qdcsm4}. At last the matrix elements of 5-body Hamiltonian
can be obtained,
\begin{equation}
\langle\Phi_{{\alpha}K}|H_5|\Phi_{{\alpha}K^{\prime}}\rangle
=6\langle\Phi_{{\alpha}K}|H_{12}|\Phi_{{\alpha}K^{\prime}}\rangle
+4\langle\Phi_{{\alpha}K}|H_{15}|\Phi_{{\alpha}K^{\prime}}\rangle
\label{zjz}
\end{equation}
The matrix elements of five body Hamiltonian on the physical bases
can be obtained from the matrix elements on the symmetry bases and
the transformation coefficients.

\section{Quark models and calculation method}

\subsection{Naive quark cluster model}
The Hamiltonian of naive quark cluster model is \cite{Isgur},
\begin{equation}
H=\sum (m_i+\frac{p_i^2}{2m_i}) -T_{CM} + \sum_{i>j=1}^5
(V^C_{ij}+V^G_{ij}),
\end{equation}

\begin{equation}
    T_{CM}=\frac{1}{2M}\left( \sum^5_{i=1}\vec{p}_i
    \right)^2,~~~M=\sum_{i=1}^5m_i,
\end{equation}

\begin{equation}
V_{ij}^{G}=\alpha_s\frac{\vec{\lambda}_i \cdot \vec{\lambda}_j}{4}
\left[\frac{1}{r_{ij}}-\frac{\pi\delta(\vec{~r})}{2}\left(\frac{1}{m_i^2}+\frac{1}
{m_j^2}+\frac{4\vec{\sigma}_i\cdot\vec{\sigma}_j}{3m_im_j}\right)\right],
\label{oge}
\end{equation}

\begin{equation}
V_{ij}^{C}=-\alpha_c\vec{\lambda}_i \cdot \vec{\lambda}_j r_{ij}^2.
\label{conf}
\end{equation}
All of the symbols retain their original meaning as in the
Ref.(\cite{Isgur}).

The single particle orbital wavefunction in the naive quark model is
as follows:
\begin{equation}
\phi_L(\vec{r})=(\frac{1}{\pi
b^2})^{\frac{3}{4}}e^{-\frac{1}{2b^2}(\vec{r}+\frac{\vec{S}}{2})^2},
\label{spol}
\end{equation}
\begin{equation}
\phi_R(\vec{r})=(\frac{1}{\pi
b^2})^{\frac{3}{4}}e^{-\frac{1}{2b^2}(\vec{r}-\frac{\vec{S}}{2})^2},
\label{spor}
\end{equation}
\begin{equation}
\phi_U(\vec{r})=(\frac{1}{\pi
b^2})^{\frac{3}{4}}e^{-\frac{1}{2b^2}(\vec{r}-\vec{T})^2},
\label{spou}
\end{equation}
where $\frac{\vec{S}}{2}$ and $\vec{T}$ are the coordinates of the
reference center of each quark cluster.

\subsection{Chiral quark model}
We choose the Salamanca model as representative of this class of
models. Details of the model can be found in Ref.\cite{chiral}. Here
we display only the Hamiltonian,
\begin{equation}
H=\sum (m_i+\frac{p_i^2}{2m_i}) + \sum_{i>j=1}^5
(V^C_{ij}+V^G_{ij}+V^\chi_{ij}+V^{\sigma}_{ij}),
\end{equation}

\begin{eqnarray}
V_{ij}^{\chi} &=& \frac{1}{3}\alpha_{ch}
\frac{\Lambda^2}{\Lambda^2-m_{\chi}^2}m_\chi \left[
\frac{e^{-m_{\chi} r_{ij}}}{m_{\chi} r_{ij}}-
\frac{\Lambda^3}{m_{\pi}^3}\frac{e^{-m_{\Lambda}
r_{ij}}}{m_{\Lambda} r_{ij}} \right]
\vec{\sigma}_i\cdot\vec{\sigma}_j \vec{\lambda}_i \cdot \vec{\lambda}_j, ~~~\chi=\pi,K,\eta  \\
V_{ij}^{\sigma} &=& -\alpha_{ch} \frac{4m_q^2}{m_\pi^2}
\frac{\Lambda^2}{\Lambda^2-m_{\sigma}^2}m_\sigma \left[
\frac{e^{-m_\sigma r_{ij}}}{m_\sigma
r_{ij}}-\frac{\Lambda}{m_\sigma}\frac{e^{-\Lambda r_{ij}}}{m_\Lambda
r_{ij}})\right],
 \nonumber
\end{eqnarray}
$V_{ij}^{\chi}$ and $V_{ij}^{\sigma}$ are pseudo scalar and scalar
meson exchange potentials. The color confinement and
one-gluon-exchange potentials and the single particle orbital
wavefunctions are the same as those in the naive quark model.

\subsection{Quark delocalization, color screening model}
The quark delocalization, color screening model (QDCSM) is an
extension of the naive quark cluster model and was developed with
the aim of addressing multi-quark systems \cite{qdcsm1,qdcsm2}.
First of all, a quark-delocalization similar to the percolation of
electrons in atoms is introduced to take into account the
contribution of orbital excitation or the mutual distortion of
hadrons in the interaction region. Second, a different
parametrization of the confinement interaction is assumed for the
quark pairs in different states. The parametrization is an effort to
account for the QCD interactions corresponding to various hidden
color configurations in the multi-quark system which have not been
modeled in two body interaction model. The main advantage of QDCSM
is that it allows the multi-quark system to choose its most
favorable configuration through its own dynamics. This is
accomplished by varying the energy of the system with respect to the
delocalization parameter, which is a tentative approach to take into
account the self-consistency of the quark and gluon distributions in
the course of hadron interaction process.

This model reproduces the existing baryon-baryon interaction data
well \cite{qdcsm1,qdcsm2} (bound-state deuteron as well as $NN$,
$N\Lambda$ and $N\Sigma$ scattering). It is therefore interesting to
apply the QDCSM to study the pentaquark system. Some generalizations
are needed here: the quark can delocalize among clusters and the
color confinement between quarks in different clusters is screened
as before, but now the clusters may be colorful as well as
colorless. We admit there should be difference of QCD vacuum between
colorless hadrons and colorful ones and so color screening should be
different but in order to avoid new parameters we employ the
original one tentatively.

The quark delocalization is realized by replacing the single
particle orbital wavefunctions ($\phi_L$ and $\phi_R$ by $\psi_l$
and $\psi_r$):
\begin{eqnarray}
\psi_l&=&(\phi_L+\epsilon_1\phi_R+\epsilon_2  \phi_U)/N_l, \nonumber\\
\psi_r&=&(\phi_R+\epsilon_1\phi_L+\epsilon_2 \phi_U)/N_r,\\
\psi_u&=&(\phi_U+\epsilon_3\phi_L+\epsilon_3\phi_R)/N_u,
\end{eqnarray}
\begin{eqnarray}
N_l&=&\sqrt{1+\epsilon^2_1+\epsilon^2_2+2\epsilon_1\langle\phi_L|\phi_R\rangle+2\epsilon_2
\langle\phi_L|\phi_U\rangle+2\epsilon_1\epsilon_2\langle\phi_R|\phi_U\rangle},  \nonumber  \\
N_r&=&\sqrt{1+\epsilon^2_1+\epsilon^2_2+2\epsilon_1\langle\phi_L|\phi_R\rangle+2\epsilon_2
\langle\phi_R|\phi_U\rangle+2\epsilon_1\epsilon_2\langle\phi_L|\phi_U\rangle},    \\
N_u&=&\sqrt{1+2\epsilon^2_3+2\epsilon_3\langle\phi_U|\phi_L\rangle+2\epsilon_3
\langle\phi_U|\phi_R\rangle+2\epsilon^2_3\langle\phi_L|\phi_R\rangle},
\end{eqnarray}
here, $\epsilon_{1}$, $\epsilon_{2}$ and $\epsilon_{3}$ are
variational parameters determined by the dynamics of the multi-quark
system rather than adjustable (fitting) parameters.

The color screeening is realized by re-parameterizing the color
confinement interaction as follows:
\begin{equation}
V_{ij}^{C}=\left \{
\begin{array}{ll}
\mbox{$-\alpha_c\vec{\lambda}_i \cdot \vec{\lambda}_j r_{ij}^2$
\ {} \ {} \ {} \ {} \ {} if $i, j$ occur in orbits with same reference center}\\
\mbox{$-\alpha_c\vec{\lambda}_i\cdot\vec{\lambda}_j\frac{1-e^{-\mu
r_{ij}^2}}{\mu}$ \ {} \ {}if $i, j$ occur in orbits with different
reference centers}
\end{array}
\right.
\end{equation}
Details of the model can be found in Ref.\cite{qdcsm2}.

The adiabatic approximation is used here to do a systematic study.
For each given separations $S$ and $T$, the energy of 5-quark state
is calculated. (For QDCSM, the energy for given $S$ and $T$ is
obtained by varying the energy with the delocalization parameters.)
If there exists minimum energy at finite separations $S_0$ and
$T_0$, then the energy $E(S_0,T_0)$ is taken as the mass of the
state.

\section{Results and discussions}

The calculated transformation coefficients are listed in Table A1 of
appendix. (The index of diquark cluster is given in Table A2.) In
this table, all the channels are included, so it can be used to
expand the physical bases in terms of the symmetry bases, and vice
versa. This table can be used for any quark model Hamiltonian but
restricted to JW $qq-qq-\bar{q}$ cluster configuration. For
meson-baryon cluster configuration one needs other transformation
coefficients which will be given in a company paper.

All possible states within the $u$, $d$, $s$ three-flavor world have
been calculated. Both single-channel and channel coupling
calculations have been carried out with three quark models: the
naive quark model, the chiral quark model and the QDCSM. The results
are listed in Table 1. Because the parity is a good quantum number,
so the results for positive and negative parity states are listed
separately. To save space, only the lowest single channel (sc) and
channel coupling (cc) results are given. All the states with $Y = 2$
but only several states with $Y \neq 2$ are given. Because most
states with $Y \neq 2$ have quite similar features to the ones with
$Y = 2$ part. Some general features are listed below.

(1) The parity of the lowest channel is negative in all three quark
models, which is different from Jaffe-Wilczek's estimation. The
diquark with orbital, color, spin, flavor symmetry: [2], [11], [11],
[11] does have the lowest energy under the color-magnetic
interaction, however the Pauli principle excludes the $S$-wave
orbital motion between two such diquarks. The $P$-wave excitation
and the ``residue" interaction between two diquarks make the energy
of JW state higher than the state where two diquarks have the
symmetry: $[2]\times [2]$ (orbital), $[11]\times [2]$ (color),
$[11]\times [2]$ (spin) and $[11]\times [11]$ (flavor), where
$S$-wave orbital is permitted.

\begin{center}
\epsfxsize=2.90in \epsfbox{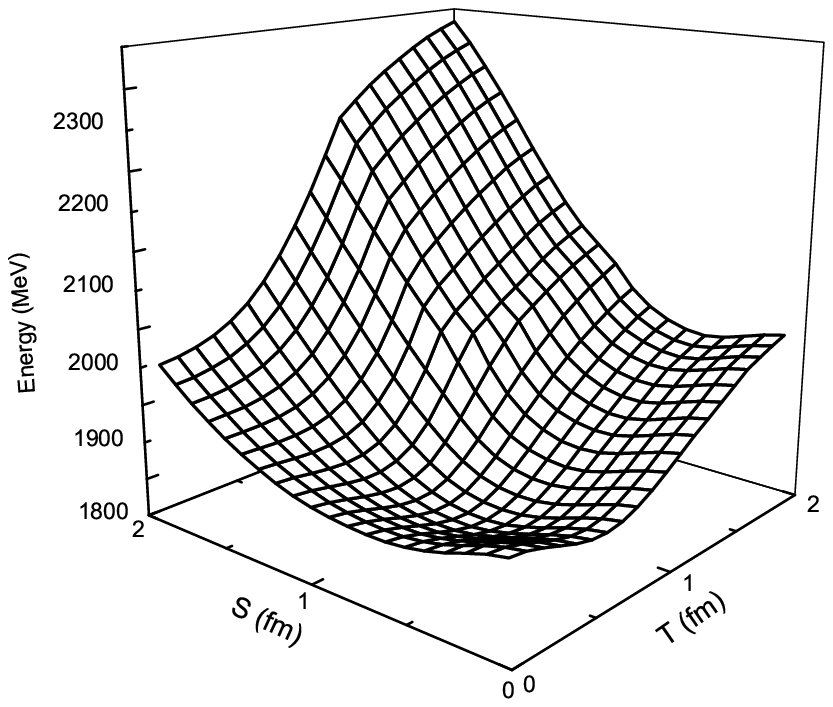}

Fig.4 The effective potential for $YIJ=20\frac{1}{2}$ with
channel-coupling in QDCSM.

\end{center}

(2) Generally there exist effective attractions for both positive
and negative parity states, thus forming resonance in both parity
states is possible. This is due to the hidden color structure of JW
configuration as discussed in the introduction part. However in most
cases, the attraction is not enough to make the energy lower than
the corresponding threshold, i.e., the sum of corresponding baryon
and meson mass. Therefore there will be no narrow resonances except
there is special mechanism to prevent the decay. Fig.4 gives the
effective potential for the channel-coupling result of $\Theta^{+}$
in QDCSM. The other two models have similar effective potentials as
QDCSM. Other states have similar effective potentials as that of
$\Theta^{+}$ shown in Fig.4. In QDCSM, the mass of $\Theta^{+}$ is
1786 MeV which is too high to match the experimentally claimed 1540
MeV. Other two models give even higher mass of $\Theta^+$. For
$\Xi^{--}$, the QDCSM mass is about 1884 MeV, a little higher than
the claimed 1862 MeV. Other two models' mass of $\Xi^{--}$ are much
higher (1974 MeV for chiral quark model and 2145 for naive quark
model).

(3) The spectroscopy of QDCSM and chiral quark model are quite
similar except a mass shift. This result is quite consistent with
our former work on $NN$ interaction \cite{NN}, where the two models
with quite different mechanism of intermediate range attraction give
similar results of deuteron properties and $NN$ scattering phase
shifts, and shows the $\sigma$ meson effect can be replaced by QDCSM
mechanism for pentaquark systems as well. However the naive qauark
model results are quite different. The naive quark model is quite
possible not a realistic one for multi-quark system even though it
is a good model for single hadrons because it can not reproduce the
intermediate range $NN$ attraction and the deuteron properties.

(4) In QDCSM, the majority of the lowest energy states have the
triangle pattern: the separations between the two diquarks are $S =
0.8 \sim 1.5$ fm and the separation between the anti-quark and the
center of the two diquarks is $T = 0.3 \sim 0.9$ fm except few
positive parity states. While, in the chiral quark model and naive
quark model, the lowest energy states always have the linear
pattern: the separations between the two diquarks are $S = 0.3 \sim
0.8$ fm and the separation between the anti-quark to the diquark
center is 0.1fm. We don't have any experimental data to check which
one is more realistic, lattice QCD calculation might provide useful
information.

\begin{center}
Table 1. The mass of the pentaquark in various quark models (S,T in
fm and E in MeV).

\begin{tabular}{|ccc|c|cccccc|ccc|ccc|}
\hline \multicolumn{4}{|c|}{} & \multicolumn{6}{|c|}{QDCSM} &
\multicolumn{3}{|c|}{Chiral
Quark Model} & \multicolumn{3}{|c|}{Naive Quark Model} \\
\hline $Y$ & $I$ & $J^{P}$ & & ~~$E$~~ & ~$S$~ & ~$T$~ &
~$\epsilon_1$~ & ~$\epsilon_2$~ & ~$\epsilon_3$~ & ~~~~$E$~~~~ &
~~$S$~~~ & $T$ & ~~~~$E$~~~~ & ~~$S$~~~ & $T$\\ \hline 2 & 2 &
$\frac{5}{2}^{+}$ & sc & 2150 & 1.4 &
0.6 & 0.8 & 0.99 & 0.8 & 2387 & 0.7 & 0.1 & 2498 & 0.7 & 0.1\\
2 & 2 & $\frac{3}{2}^{+}$ & sc & 2113 & 1.3 &
0.6 & 0.9 & 0.99 & 0.99 & 2338 & 0.6 & 0.1 & 2463 & 0.7 & 0.1\\
 & & & cc & 2107 & 1.3 & 0.6 & 0.9 & 0.99 & 0.99 & 2330 & 0.6
& 0.1 & 2453 & 0.7 & 0.1\\
2 & 2 & $\frac{3}{2}^{-}$ & sc & 2040 & 1.0 &
0.9 & 0.4 & 0.7 & 0.99 & 2266 & 0.5 & 0.1 & 2342 & 0.5 & 0.1\\
 & & & cc & 2038 & 1.0 & 0.9 & 0.2 & 0.6 & 0.99 & 2244 & 0.6
& 0.1 & 2316 & 0.6 & 0.1\\
2 & 2 & $\frac{1}{2}^{+}$ & sc & 2255 & 1.2 &
0.8 & 0.1 & 0.5 & 0.9 & 2495 & 0.8 & 0.1 & 2569 & 0.8 & 0.1\\
 & & & cc & 2246 & 1.2 & 0.7 & 0.1 & 0.6 & 0.8 & 2484 & 0.8
& 0.1 & 2556 & 0.7 & 0.1\\
2 & 2 & $\frac{1}{2}^{-}$ & sc & 2082 & 1.0 &
0.9 & 0.4 & 0.7 & 0.99 & 2329 & 0.6 & 0.1 & 2416 & 0.6 & 0.1\\
 & & & cc & 2080 & 1.0 & 0.9 & 0.1 & 0.5 & 0.99 & 2311 & 0.7
& 0.1 & 2393 & 0.7 & 0.1\\
2 & 1 & $\frac{5}{2}^{+}$ & sc & 2245 & 1.1 &
0.8 & 0.9 & 0.99 & 0.9 & 2519 & 0.5 & 0.1 & 2608 & 0.5 & 0.1\\
2 & 1 & $\frac{5}{2}^{-}$ & sc & 2011 & 1.0 &
0.8 & 0.5 & 0.8 & 0.99 & 2223 & 0.4 & 0.1 & 2321 & 0.4 & 0.1\\
 & & & cc & 2009 & 1.0 & 0.8 & 0.2 & 0.6 & 0.99 & 2206 & 0.6
& 0.1 & 2298 & 0.6 & 0.1\\
2 & 1 & $\frac{3}{2}^{+}$ & sc & 2037 & 1.2 &
0.6 & 0.3 & 0.6 & 0.9 & 2209 & 0.6 & 0.1 & 2405 & 0.7 & 0.1\\
 & & & cc & 1977 & 1.3 & 0.3 & 0.7 & 0.99 & 0.6 & 2148 & 0.5
& 0.1 & 2377 & 0.6 & 0.1\\
2 & 1 & $\frac{3}{2}^{-}$ & sc & 1909 & 0.9 &
0.8 & 0.3 & 0.6 & 0.99 & 2061 & 0.4 & 0.1 & 2228 & 0.4 & 0.1\\
 & & & cc & 1882 & 0.8 & 0.8 & 0.1 & 0.5 & 0.99 & 2014 & 0.5
& 0.1 & 2157 & 0.5 & 0.1\\
2 & 1 & $\frac{1}{2}^{+}$ & sc & 2010 & 1.2 &
0.5 & 0.3 & 0.6 & 0.9 & 2172 & 0.6 & 0.1 & 2377 & 0.7 & 0.1\\
 & & & cc & 1931 & 1.2 & 0.3 & 0.8 & 0.99 & 0.8 & 2089 & 0.4
& 0.1 & 2331 & 0.6 & 0.1\\
2 & 1 & $\frac{1}{2}^{-}$ & sc & 1886 & 0.9 &
0.8 & 0.6 & 0.8 & 0.99 & 2034 & 0.3 & 0.1 & 2231 & 0.4 & 0.1\\
 & & & cc & 1868 & 0.8 & 0.8 & 0.1 & 0.5 & 0.99 & 2000 & 0.4
& 0.1 & 2181 & 0.5 & 0.1\\
2 & 0 & $\frac{5}{2}^{+}$ & sc & 2242 & 1.2 &
0.7 & 0.1 & 0.6 & 0.8 & 2483 & 0.7 & 0.1 & 2557 & 0.7 & 0.1\\
2 & 0 & $\frac{3}{2}^{+}$ & sc & 2117 & 1.0 &
0.7 & 0.9 & 0.8 & 0.99 & 2326 & 0.4 & 0.1 & 2488 & 0.7 & 0.1\\
 & & & cc & 2079 & 1.1 & 0.6 & 0.8 & 0.99 & 0.9 & 2273 & 0.3
& 0.1 & 2451 & 0.5 & 0.1\\
2 & 0 & $\frac{3}{2}^{-}$ & sc & 1871 & 0.8 &
0.8 & 0.3 & 0.6 & 0.99 & 2011 & 0.3 & 0.1 & 2219 & 0.4 & 0.1\\
 & & & cc & 1870 & 0.8 & 0.8 & 0.3 & 0.6 & 0.99 & 2007 & 0.4
& 0.1 & 2207 & 0.5 & 0.1\\
2 & 0 & $\frac{1}{2}^{+}$ & sc & 1915 & 1.1 &
0.5 & 0.1 & 0.4 & 0.9 & 2054 & 0.7 & 0.1 & 2316 & 0.7 & 0.1\\
 & & & cc & 1868 & 1.2 & 0.1 & 0.5 & 0.99 & 0.5 & 1999 & 0.5
& 0.1 & 2280 & 0.6 & 0.1\\
2 & 0 & $\frac{1}{2}^{-}$ & sc & 1787 & 0.8 &
0.7 & 0.3 & 0.6 & 0.99 & 1887 & 0.3 & 0.1 & 2109 & 0.3 & 0.1\\
 & & & cc & 1786 & 0.8 & 0.7 & 0.1 & 0.5 & 0.99 & 1886 & 0.3
& 0.1 & 2100 & 0.5 & 0.1\\ \hline

\end{tabular}
\end{center}

\begin{center}
\begin{tabular}{|ccc|c|cccccc|ccc|ccc|}
\hline \multicolumn{4}{|c|}{} & \multicolumn{6}{|c|}{QDCSM} &
\multicolumn{3}{|c|}{Chiral
Quark Model} & \multicolumn{3}{|c|}{Naive Quark Model} \\
\hline $Y$ & $I$ & $J^{P}$ & & ~~$E$~~ & ~$S$~ & ~$T$~ &
~$\epsilon_1$~ & ~$\epsilon_2$~ & ~$\epsilon_3$~ & ~~~~$E$~~~~ &
~~$S$~~~ & $T$ & ~~~~$E$~~~~ & ~~$S$~~~ & $T$\\ \hline 1 &
$\frac{5}{2}$ & $\frac{5}{2}^{+}$ & sc & 2103 & 1.4 & 0.6 &
0.9 & 0.99 & 0.7 & 2366 & 0.7 & 0.1 & 2457 & 0.7 & 0.1\\
1 & $\frac{3}{2}$ & $\frac{5}{2}^{+}$ & sc & 2041 & 1.4 & 0.5 &
0.7 & 0.99 & 0.6 & 2273 & 0.6 & 0.1 & 2457 & 0.7 & 0.1\\
 & & & cc & 2040 & 1.4 & 0.5 &
0.7 & 0.99 & 0.6 & 2271 & 0.6 & 0.1 & 2457 & 0.7 & 0.1\\
1 & $\frac{1}{2}$ & $\frac{5}{2}^{+}$ & sc & 2150 & 1.2 & 0.7 &
0.8 & 0.99 & 0.9 & 2424 & 0.5 & 0.1 & 2531 & 0.8 & 0.1\\
 & & & cc & 2149 & 1.2 & 0.7 &
0.7 & 0.99 & 0.8 & 2421 & 0.5 & 0.1 & 2531 & 0.8 & 0.1\\
1 & $\frac{1}{2}$ & $\frac{5}{2}^{-}$ & sc & 1906 & 1.0 &
0.8 & 0.5 & 0.8 & 0.99 & 2123 & 0.4 & 0.1 & 2289 & 0.4 & 0.1\\
 & & & cc & 1903 & 1.0 & 0.8 & 0.2 & 0.6 & 0.99 & 2101 & 0.6
& 0.1 & 2260 & 0.6 & 0.1\\
0 & 2 & $\frac{5}{2}^{+}$ & sc & 2161 & 1.4 & 0.6 &
0.8 & 0.99 & 0.7 & 2400 & 0.7 & 0.1 & 2483 & 0.7 & 0.1\\
 & & & cc & 2161 & 1.4 & 0.6 &
0.8 & 0.99 & 0.7 & 2400 & 0.7 & 0.1 & 2483 & 0.7 & 0.1\\
0 & 2 & $\frac{5}{2}^{-}$ & sc & 2002 & 1.0 & 0.8 &
0.5 & 0.8 & 0.99 & 2211 & 0.4 & 0.1 & 2319 & 0.4 & 0.1\\
 & & & cc & 2001 & 1.0 & 0.8 &
0.2 & 0.6 & 0.99 & 2196 & 0.6 & 0.1 & 2296 & 0.6 & 0.1\\
0 & 1 & $\frac{5}{2}^{+}$ & sc & 2110 & 1.4 & 0.5 &
0.7 & 0.99 & 0.6 & 2323 & 0.6 & 0.1 & 2483 & 0.7 & 0.1\\
 & & & cc & 2109 & 1.4 & 0.5 &
0.7 & 0.99 & 0.6 & 2319 & 0.6 & 0.1 & 2483 & 0.7 & 0.1\\
0 & 1 & $\frac{5}{2}^{-}$ & sc & 1944 & 0.9 & 0.8 &
0.5 & 0.7 & 0.99 & 2120 & 0.4 & 0.1 & 2319 & 0.4 & 0.1\\
 & & & cc & 1941 & 0.9 & 0.8 &
0.2 & 0.6 & 0.99 & 2106 & 0.5 & 0.1 & 2296 & 0.6 & 0.1\\
0 & 0 & $\frac{5}{2}^{+}$ & sc & 2215 & 1.2 & 0.7 &
0.1 & 0.6 & 0.8 & 2444 & 0.7 & 0.1 & 2564 & 0.7 & 0.1\\
 & & & cc & 2211 & 1.2 & 0.7 &
0.2 & 0.7 & 0.8 & 2440 & 0.7 & 0.1 & 2563 & 0.7 & 0.1\\
0 & 0 & $\frac{5}{2}^{-}$ & sc & 1985 & 1.0 & 0.8 &
0.5 & 0.8 & 0.99 & 2181 & 0.4 & 0.1 & 2321 & 0.4 & 0.1\\
 & & & cc & 1983 & 0.9 & 0.8 &
0.2 & 0.6 & 0.99 & 2164 & 0.5 & 0.1 & 2301 & 0.6 & 0.1\\
-1 & $\frac{3}{2}$ & $\frac{1}{2}^{+}$ & sc & 2109 & 1.1 &
0.5 & 0.1 & 0.5 & 0.99 & 2258 & 0.7 & 0.1 & 2404 & 0.7 & 0.1\\
 & & & cc & 2036 & 1.2 & 0.1 & 0.7 & 0.99 & 0.5 & 2179 & 0.5
& 0.1 & 2343 & 0.6 & 0.1\\
-1 & $\frac{3}{2}$ & $\frac{1}{2}^{-}$ & sc & 1895 & 0.7 &
0.7 & 0.3 & 0.6 & 0.99 & 1987 & 0.2 & 0.1 & 2158 & 0.3 & 0.1\\
 & & & cc & 1884 & 0.7 & 0.7 & 0.0 & 0.5 & 0.99 & 1974 & 0.3
& 0.1 & 2145 & 0.4 & 0.1\\
-1 & $\frac{1}{2}$ & $\frac{5}{2}^{+}$ & sc & 2211 & 1.2 & 0.6 &
0.5 & 0.8 & 0.8 & 2405 & 0.6 & 0.1 & 2539 & 0.7 & 0.1\\
 & & & cc & 2175 & 1.5 & 0.1 &
0.6 & 0.99 & 0.3 & 2367 & 0.6 & 0.1 & 2510 & 0.7 & 0.1\\
-2 & 1 & $\frac{5}{2}^{+}$ & sc & 2272 & 1.4 & 0.4 &
0.8 & 0.99 & 0.5 & 2462 & 0.6 & 0.1 & 2539 & 0.7 & 0.1\\
 & & & cc & 2272 & 1.4 & 0.4 &
0.8 & 0.99 & 0.5 & 2461 & 0.6 & 0.1 & 2539 & 0.7 & 0.1\\
-2 & 0 & $\frac{5}{2}^{+}$ & sc & 2244 & 1.5 & 0.1 &
0.6 & 0.99 & 0.3 & 2423 & 0.6 & 0.1 & 2539 & 0.7 & 0.1\\
 & & & cc & 2241 & 1.5 & 0.1 &
0.6 & 0.99 & 0.3 & 2416 & 0.6 & 0.1 & 2539 & 0.7 & 0.1\\
-3 & $\frac{1}{2}$ & $\frac{5}{2}^{+}$ & sc & 2323 & 1.4 & 0.1 & 0.7
& 0.99 & 0.3 & 2491 & 0.5 & 0.1 & 2568 & 0.6 & 0.1\\ \hline

\end{tabular}
\end{center}

There were discussions on the systematics of the states in
anti-decuplet, 27-plet and 35-plet
\cite{JW,10plet,27plet,35plet,Bijker}. These states are classified
by flavor symmetry. Here, we only consider flavor symmetry of
4-quark system, when coupled to anti-quark, the states in $1_{f}$,
$8_{f}$, $10_{f}$, $\bar{10}_{f}$, $27_{f}$, $35_{f}$ will be
mixed, i.e., $[4]\otimes[11]\rightarrow[51]\oplus[411]$;
$[31]\otimes[11]\rightarrow[42]\oplus[411]\oplus[321]$;
$[22]\otimes[11]\rightarrow[33]\oplus[321]$;
$[211]\otimes[11]\rightarrow[321]\oplus[222]$. So the states we
discuss here are mixed ones. In the following we give the main
features of these states in our calculation and compare them with
other model results.

\begin{figure}
\begin{center}
\epsfxsize=2.5in \epsfbox{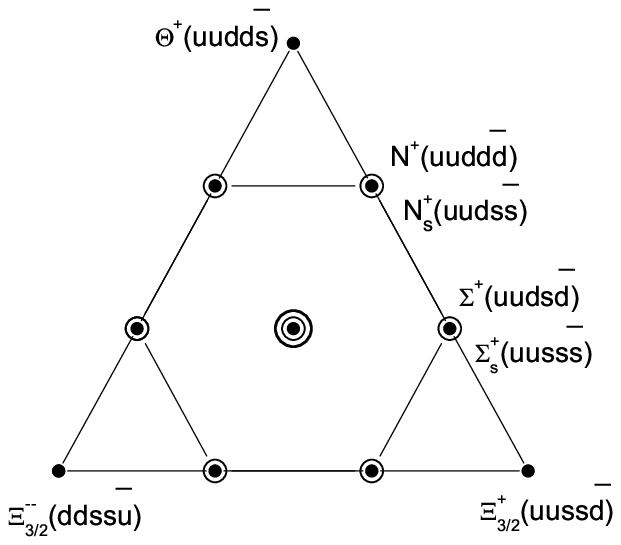}

Fig.5(a) Quark content of representative members of the $8_f +
\overline{10}_{f}$ states.
\end{center}
\end{figure}
\begin{figure}
\begin{center}
\epsfysize=3.0in \epsfbox{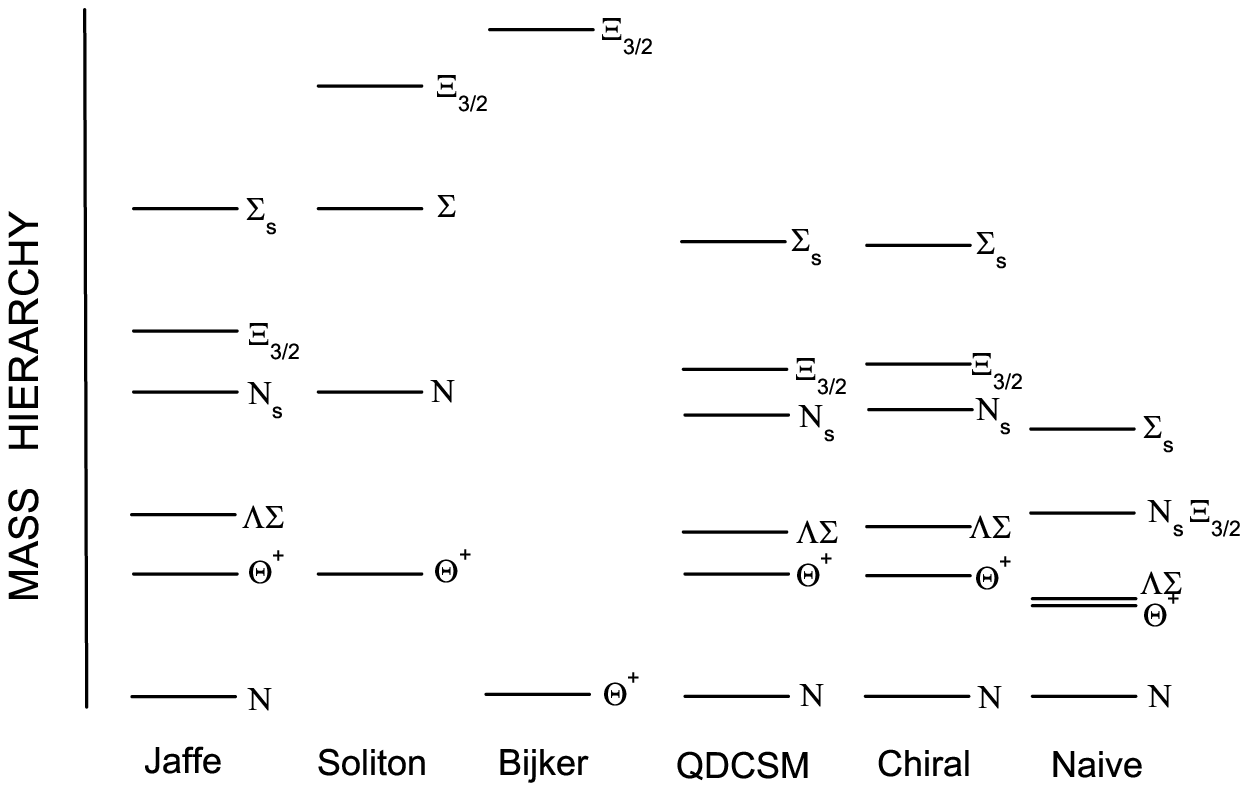}

Fig.5(b) Relative masses of states in the $8_f +
\overline{10}_{f}$ with $J^{P} = \frac{1}{2}^{+}$ in three quark
models, compared with the mass hierarchy in Ref.\cite{JW} and
Ref.\cite{Bijker}.
\end{center}
\end{figure}

(1) $8_f + \overline{10}_{f}$ states: The quark contents of
representative states in the $8_f + \overline{10}_{f}$ are given in
Fig.5(a). The mass spectrum of the states with
$J^{P}=\frac{1}{2}^{+}$ in three quark models are summarized in Fig.
5(b), and compared with Jaffe-Wilczek's and chiral soliton model
results \cite{JW}. Obviously, the order of the states of JW, chiral
quark model and QDCSM is similar, the chiral soliton model and the
naive quark model results are different from the above three. In
quark model, generally the states with more $s$-quarks lie higher,
so $\Sigma_s$ with three strange quarks has the highest energy,
while $N$ without strange quark has the lowest energy, $N_s$,
$\Xi_{3/2}$ (with two strange quarks) and $\Lambda, \Sigma,
\Theta^+$ (with one strange quark) lie between. While in chiral
soliton model the $\Theta^+$ is the lowest state, where flavor
$SU(3)$ symmetry is implied. However the flavor $SU(3)$ symmetry is
broken by the large strange quark mass. The mass spectrum of the
states with $J^{P}=\frac{1}{2}^{-}$ in three quark models are
summarized in Fig. 5(c), and compared with chiral soliton models
\cite{10plet}. The results of Bijker's \cite{Bijker} are also
listed. For $J^{P}=\frac{1}{2}^{-}$, the order of the states is
different from the one for $J^{P}=\frac{1}{2}^{+}$, but all the
three quark models are similar. The chiral soliton model results are
quite different from that of three quark models. Other states with
$J^{P}=\frac{3}{2}^{\pm}$ and $J^{P}=\frac{5}{2}^{\pm}$ are also
calculated. To save space, we have not listed all the results here.

\begin{figure}
\begin{center}
\epsfysize=3.4in \epsfbox{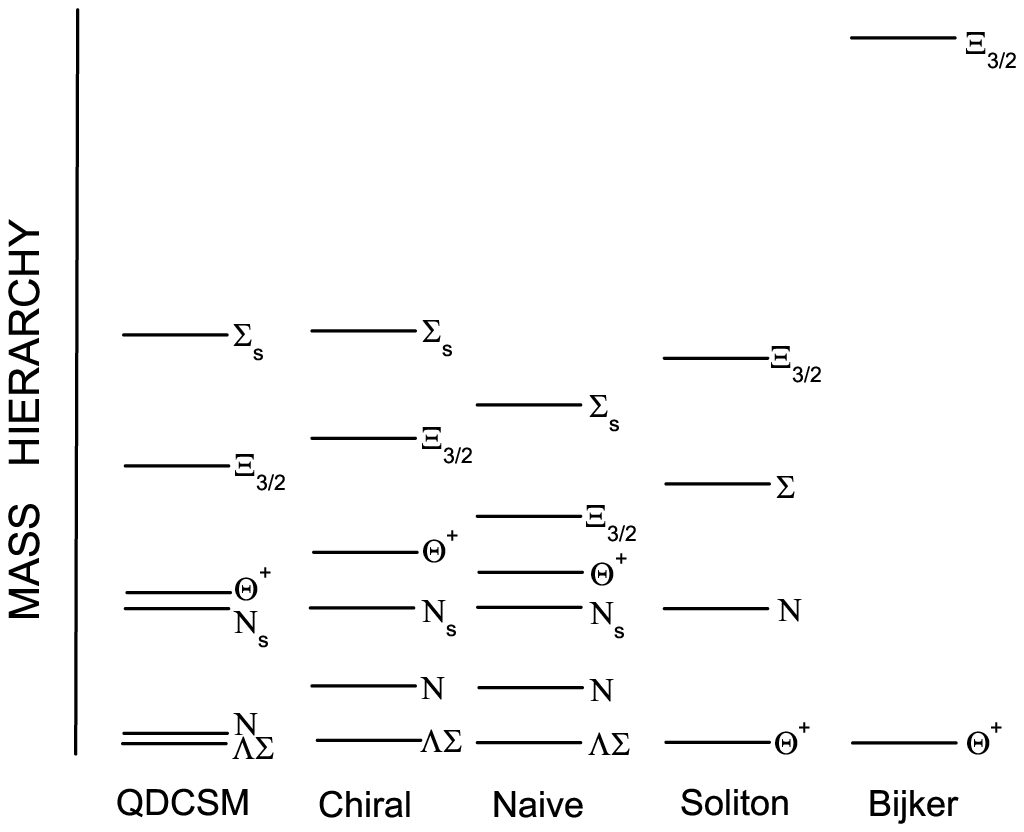}

Fig.5(c) Relative masses of states in the $8_f +
\overline{10}_{f}$ with $J^{P} = \frac{1}{2}^{-}$ in three quark
models, compared with the mass hierarchy in Ref.\cite{10plet} and
Ref.\cite{Bijker}.
\end{center}
\end{figure}

(2) 27-plet: The quark contents of representative states in the
27-plet are given in Fig.6(a). The mass spectrum of the states with
$J^{P}=\frac{3}{2}^{\pm}$ in three quark models are summarized in
Fig. 6(b-c), and compared with the chiral soliton model
\cite{27plet} and Bijker's results \cite{Bijker}. We see again that
the chiral quark model and QDCSM results are similar but quite
different from the results of chiral soliton model and Bijker's
results.

\begin{figure}
\begin{center}
\epsfxsize=3.0in \epsfbox{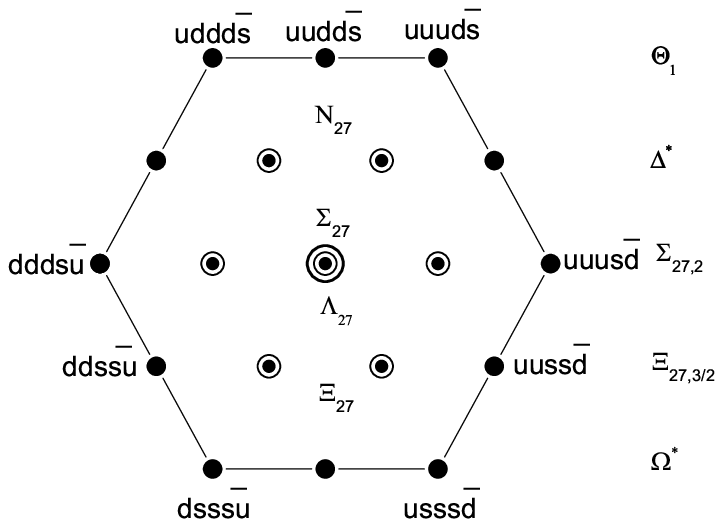}

Fig.6(a) Quark content of representative members of the 27-plet
states.
\end{center}
\end{figure}

\begin{figure}
\begin{center}
\epsfysize=3.2in \epsfbox{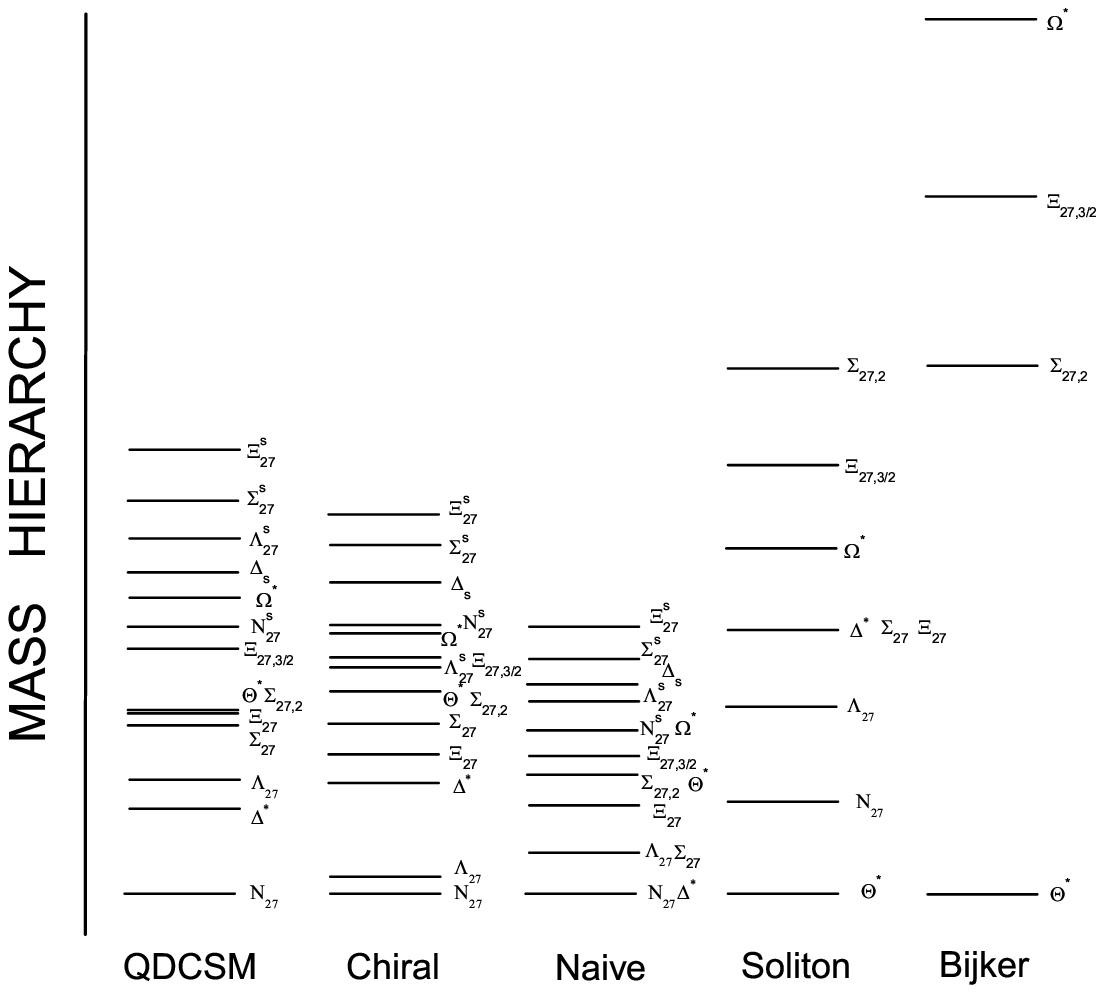}

Fig.6(b) Relative masses of states in the 27-plet with $J^{P} =
\frac{3}{2}^{-}$ in three quark models, compared with the mass in
Ref.\cite{27plet} and Ref.\cite{Bijker}.
\end{center}
\end{figure}

\begin{figure}
\begin{center}
\epsfysize=3.2in \epsfbox{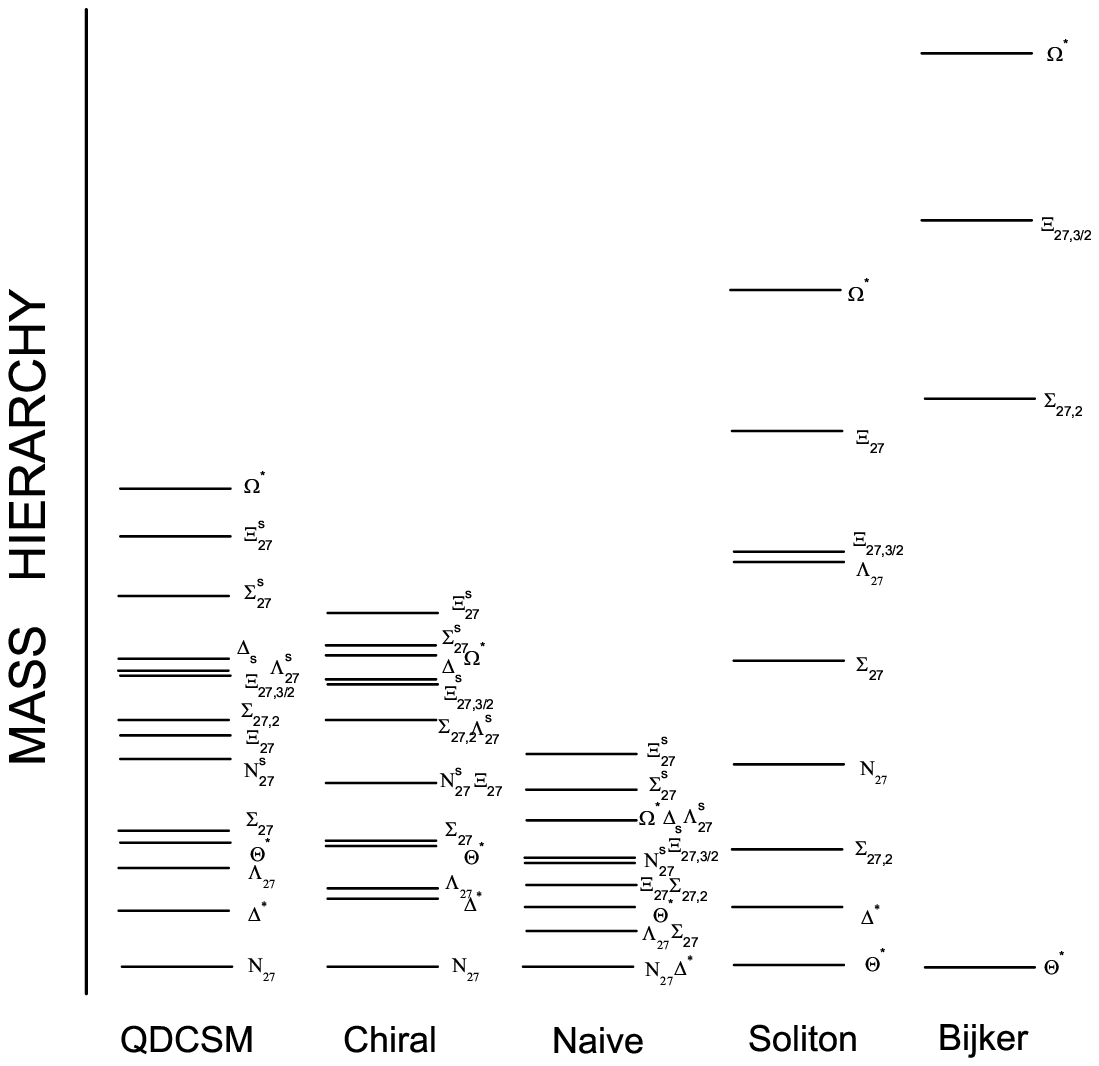}

Fig.6(c) Relative masses of states in the 27-plet with $J^{P} =
\frac{3}{2}^{+}$ in three quark models, compared with the mass in
Ref.\cite{27plet} and Ref.\cite{Bijker}.
\end{center}
\end{figure}

(3) 35-plet: The quark contents of representative states in the
35-plet are given in Fig.7(a). The mass spectrum of the states with
$J^{P}=\frac{5}{2}^{+}$ in three quark models are summarized in Fig.
7(b), and compared with the chiral soliton model \cite{35plet} and
Bijker's results \cite{Bijker}. One sees once more that the chiral
quark model and QDCSM results are similar but different from the
chiral soliton model and Bijker's results.
\begin{figure}
\begin{center}
\epsfxsize=3.0in \epsfbox{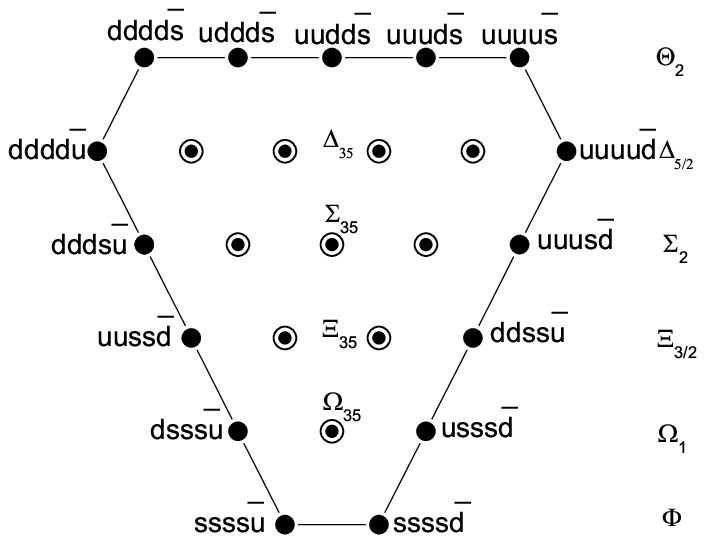}

Fig.7(a) Quark content of representative members of the 35-plet
states.
\end{center}
\end{figure}

\begin{figure}
\begin{center}
\epsfysize=3.0in \epsfbox{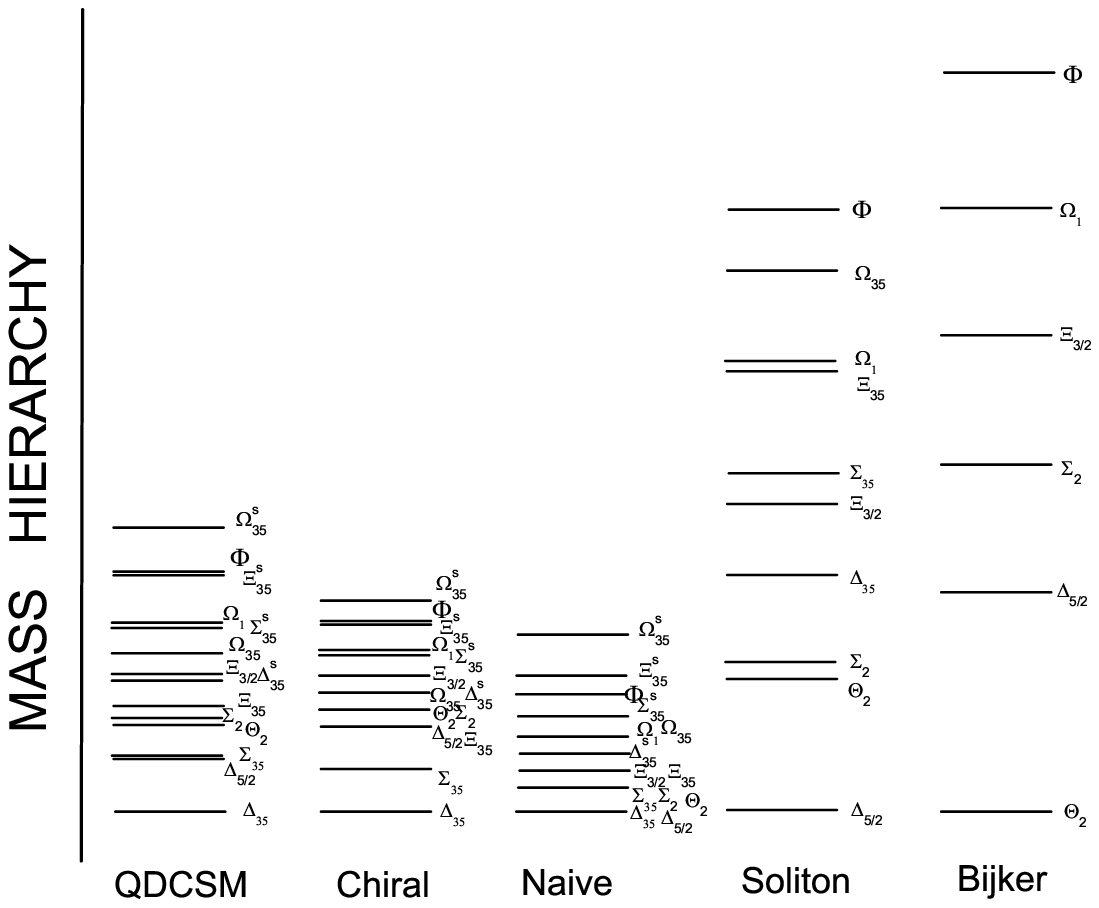}

Fig.7(b) Relative masses of states in the 35-plet with $J^{P} =
\frac{5}{2}^{+}$ in three quark models, compared with the mass in
Ref.\cite{35plet} and Ref.\cite{Bijker}.
\end{center}
\end{figure}

\section{Summary}
The dibaryon states had been claimed and disappeared few times, the
pentaquark $\Theta^{+}$ might disappear and the tetra-quark states
might be not confirmed. However multi quark state search will be
continued. Multi quark components in meson and baryon can not be
denied based on QCD. Meson-meson, meson-baryon, baryon-baryon,
baryon-antibaryon scattering are measured. If one wants to
understand all of these physics from quark-gluon degree of freedom
one needs few-body calculation method and the group theoretic method
is one of the powerful ones. This paper reports the needed group
theory results for five-quark calculation within Jaffe-Wilczek quark
cluster configuration. A systematic study of all possible pentaquark
states within $u,d,s$ three-flavor world in three quark models is
performed. The powerful feature of group theory method is shown by
the large amount of spectroscopy data obtained easily in this
approach.

About the physical results we emphasize first that in general there
exists effective attraction for both parity states because of the JW
hidden color configuration. So it is possible to form five-quark
resonance, because once such a state is formed it can not decay into
colorful sub-systems immediately and must transit to colorless
sub-systems through color rearrangement first then decay. This is
similar to compound nucleus formation but due to color confinement.
It is a new kind microscopic resonance, we call it color confinement
resonance. The transition rate is determined by the transition
interaction between hidden color states and colorless ones. Up to
now we don't have any idea about this transition interaction. So we
can not make any definite prediction about the width of these
resonances. Lattice QCD might study this transition interaction
which is highly expected.

Second, the chiral quark model and QDCSM give similar pentaquark
mass spectroscopy and they are different from that of chiral soliton
model. The $SU(3)$ flavor symmetry is broken by the large strange
quark mass. The similarity of pentaquark mass spectroscopy of chiral
quark model and QDCSM means the $\sigma$ meson effect can be
replaced by quark delocalization and color screening mechanism as
has been verified in $NN$ intermediate range attraction.

This work is supported by NSFC grant 90503011 and 10375030.

\vspace{0.2in}

\section*{Appendix}

 Table A1. Transformation coefficients between physical
bases and symmetry bases. The column labels are $[\nu_4]$,
$[\mu_4]$, $[f_4]$, $[J_4]$, $[I_4]$. For the first four labels, 1
stands for the symmetry label [4]; 2, [31]; 3, [22]; 4, [211], and
for the last one, 1 stands for the quantum number 2; 2,
$\frac{3}{2}$; 3, 1; 4, $\frac{1}{2}$; 5, 0. The row labels
$D_1$$D_2$ stand for two diquark clusters, the index of which are
listed in Table A2. $[J_4]$, $[I_4]$ are the same labels as the one
in column labels.

\begin{tiny}
\begin{tabular}{c|cccccccccccccc}  \hline
\multicolumn{ 4}{c}{$Y= 2$~~  $I=2          $~~  $J=\frac{5}{2}$~~
$Y_4=\frac{4}{3} $}  \\  \hline
 $D_1D_2J_4I_4$ & 21111 \\  \hline
 6~ 6~ 1~ 1 & $ 1$  \\ \hline
\end{tabular}

\begin{tabular}{c|cccccccccccccc}  \hline
\multicolumn{ 5}{c}{$Y= 2$~~  $I=2          $~~  $J=\frac{3}{2}$~~
$Y_4=\frac{4}{3} $}  \\  \hline
 $D_1D_2J_4I_4$ & 12121 & 22121 & 32121 & 21111 \\  \hline
 1~ 6~ 2~ 1 & $ \frac{  2}{   3}$ & $ 0$ & $ \frac{  1}{   3}$ & $ 0$  \\
 6~ 1~ 2~ 1 & $ 0$ & $-1$ & $ 0$ & $ 0$  \\
 6~ 6~ 1~ 1 & $ 0$ & $ 0$ & $ 0$ & $ 1$  \\
 6~ 6~ 2~ 1 & $-\frac{  1}{   3}$ & $ 0$ & $ \frac{  2}{   3}$ & $ 0$  \\ \hline
\end{tabular}

\begin{tabular}{c|cccccccccccccc}  \hline
\multicolumn{ 5}{c}{$Y= 2$~~  $I=2          $~~  $J=\frac{1}{2}$~~
$Y_4=\frac{4}{3} $}  \\  \hline
 $D_1D_2J_4I_4$ & 12121 & 22121 & 32121 & 23131 \\  \hline
 1~ 6~ 2~ 1 & $ \frac{  2}{   3}$ & $ 0$ & $ \frac{  1}{   3}$ & $ 0$  \\
 6~ 1~ 2~ 1 & $ 0$ & $-1$ & $ 0$ & $ 0$  \\
 6~ 6~ 2~ 1 & $-\frac{  1}{   3}$ & $ 0$ & $ \frac{  2}{   3}$ & $ 0$  \\
 6~ 6~ 3~ 1 & $ 0$ & $ 0$ & $ 0$ & $-1$  \\ \hline
\end{tabular}

\begin{tabular}{c|cccccccccccccc}  \hline
\multicolumn{ 4}{c}{$Y= 2$~~  $I=1          $~~  $J=\frac{5}{2}$~~
$Y_4=\frac{4}{3} $}  \\  \hline
 $D_1D_2J_4I_4$ & 12213 & 22213 & 32213 \\  \hline
 4~ 6~ 1~ 3 & $ \frac{  2}{   3}$ & $ 0$ & $ \frac{  1}{   3}$  \\
 6~ 4~ 1~ 3 & $ 0$ & $-1$ & $ 0$  \\
 6~ 6~ 1~ 3 & $-\frac{  1}{   3}$ & $ 0$ & $ \frac{  2}{   3}$  \\ \hline
\end{tabular}

\begin{tabular}{c|cccccccccccccc}  \hline
\multicolumn{11}{c}{$Y= 2$~~  $I=1          $~~  $J=\frac{3}{2}$~~
$Y_4=\frac{4}{3} $}  \\  \hline
 $D_1D_2J_4I_4$ & 12223 & 22223 & 24223 & 32223 & 34223 & 12213 & 22213 & 32213 & 21223 & 23223 \\  \hline
 1~ 6~ 2~ 3 & $ 0$ & $ \frac{  1}{   2}$ & $-\frac{  1}{   2}$ & $ 0$ & $ 0$ & $ 0$ & $ 0$ & $ 0$ & $ 0$ & $ 0$  \\
 4~ 6~ 1~ 3 & $ 0$ & $ 0$ & $ 0$ & $ 0$ & $ 0$ & $ \frac{  2}{   3}$ & $ 0$ & $ \frac{  1}{   3}$ & $ 0$ & $ 0$  \\
 4~ 6~ 2~ 3 & $ 0$ & $ \frac{  1}{   2}$ & $ \frac{  1}{   2}$ & $ 0$ & $ 0$ & $ 0$ & $ 0$ & $ 0$ & $ 0$ & $ 0$  \\
 6~ 1~ 2~ 3 & $-\frac{  1}{   3}$ & $ 0$ & $ 0$ & $-\frac{  1}{   6}$ & $-\frac{  1}{   2}$ & $ 0$ & $ 0$ & $ 0$ & $ 0$ & $ 0$  \\
 6~ 4~ 1~ 3 & $ 0$ & $ 0$ & $ 0$ & $ 0$ & $ 0$ & $ 0$ & $-1$ & $ 0$ & $ 0$ & $ 0$  \\
 6~ 4~ 2~ 3 & $-\frac{  1}{   3}$ & $ 0$ & $ 0$ & $-\frac{  1}{   6}$ & $ \frac{  1}{   2}$ & $ 0$ & $ 0$ & $ 0$ & $ 0$ & $ 0$  \\
 6~ 6~ 1~ 3 & $ 0$ & $ 0$ & $ 0$ & $ 0$ & $ 0$ & $-\frac{  1}{   3}$ & $ 0$ & $ \frac{  2}{   3}$ & $ 0$ & $ 0$  \\
 6~ 6~ 2~ 3 & $ 0$ & $ 0$ & $ 0$ & $ 0$ & $ 0$ & $ 0$ & $ 0$ & $ 0$ & $ \frac{  1}{   3}$ & $-\frac{  2}{   3}$  \\
 6~ 9~ 2~ 3 & $ 0$ & $ 0$ & $ 0$ & $ 0$ & $ 0$ & $ 0$ & $ 0$ & $ 0$ & $ \frac{  2}{   3}$ & $ \frac{  1}{   3}$  \\
 9~ 6~ 2~ 3 & $-\frac{  1}{   3}$ & $ 0$ & $ 0$ & $ \frac{  2}{   3}$ & $ 0$ & $ 0$ & $ 0$ & $ 0$ & $ 0$ & $ 0$  \\ \hline
\end{tabular}

\begin{tabular}{c|cccccccccccccc}  \hline
\multicolumn{13}{c}{$Y= 2$~~  $I=1          $~~  $J=\frac{1}{2}$~~
$Y_4=\frac{4}{3} $}  \\  \hline
 $D_1D_2J_4I_4$ & 12223 & 22223 & 24223 & 32223 & 34223 & 12233 & 22233 & 24233 & 32233 & 34233 & 21223 & 23223 \\  \hline
 1~ 6~ 2~ 3 & $ 0$ & $ \frac{  1}{   2}$ & $-\frac{  1}{   2}$ & $ 0$ & $ 0$ & $ 0$ & $ 0$ & $ 0$ & $ 0$ & $ 0$ & $ 0$ & $ 0$  \\
 1~ 9~ 3~ 3 & $ 0$ & $ 0$ & $ 0$ & $ 0$ & $ 0$ & $ \frac{  1}{   2}$ & $ 0$ & $ 0$ & $ \frac{  1}{   4}$ & $-\frac{  1}{   4}$ & $ 0$ & $ 0$  \\
 4~ 6~ 2~ 3 & $ 0$ & $ \frac{  1}{   2}$ & $ \frac{  1}{   2}$ & $ 0$ & $ 0$ & $ 0$ & $ 0$ & $ 0$ & $ 0$ & $ 0$ & $ 0$ & $ 0$  \\
 4~ 6~ 3~ 3 & $ 0$ & $ 0$ & $ 0$ & $ 0$ & $ 0$ & $-\frac{  1}{   6}$ & $ 0$ & $ 0$ & $-\frac{  1}{  12}$ & $-\frac{  3}{   4}$ & $ 0$ & $ 0$  \\
 6~ 1~ 2~ 3 & $-\frac{  1}{   3}$ & $ 0$ & $ 0$ & $-\frac{  1}{   6}$ & $-\frac{  1}{   2}$ & $ 0$ & $ 0$ & $ 0$ & $ 0$ & $ 0$ & $ 0$ & $ 0$  \\
 6~ 4~ 2~ 3 & $-\frac{  1}{   3}$ & $ 0$ & $ 0$ & $-\frac{  1}{   6}$ & $ \frac{  1}{   2}$ & $ 0$ & $ 0$ & $ 0$ & $ 0$ & $ 0$ & $ 0$ & $ 0$  \\
 6~ 4~ 3~ 3 & $ 0$ & $ 0$ & $ 0$ & $ 0$ & $ 0$ & $ 0$ & $ \frac{  1}{   4}$ & $-\frac{  3}{   4}$ & $ 0$ & $ 0$ & $ 0$ & $ 0$  \\
 6~ 6~ 2~ 3 & $ 0$ & $ 0$ & $ 0$ & $ 0$ & $ 0$ & $ 0$ & $ 0$ & $ 0$ & $ 0$ & $ 0$ & $ \frac{  1}{   3}$ & $-\frac{  2}{   3}$  \\
 6~ 6~ 3~ 3 & $ 0$ & $ 0$ & $ 0$ & $ 0$ & $ 0$ & $-\frac{  1}{   3}$ & $ 0$ & $ 0$ & $ \frac{  2}{   3}$ & $ 0$ & $ 0$ & $ 0$  \\
 6~ 9~ 2~ 3 & $ 0$ & $ 0$ & $ 0$ & $ 0$ & $ 0$ & $ 0$ & $ 0$ & $ 0$ & $ 0$ & $ 0$ & $ \frac{  2}{   3}$ & $ \frac{  1}{   3}$  \\
 9~ 1~ 3~ 3 & $ 0$ & $ 0$ & $ 0$ & $ 0$ & $ 0$ & $ 0$ & $-\frac{  3}{   4}$ & $-\frac{  1}{   4}$ & $ 0$ & $ 0$ & $ 0$ & $ 0$  \\
 9~ 6~ 2~ 3 & $-\frac{  1}{   3}$ & $ 0$ & $ 0$ & $ \frac{  2}{   3}$ & $ 0$ & $ 0$ & $ 0$ & $ 0$ & $ 0$ & $ 0$ & $ 0$ & $ 0$  \\ \hline
\end{tabular}

\begin{tabular}{c|cccccccccccccc}  \hline
\multicolumn{ 4}{c}{$Y= 2$~~  $I=0          $~~  $J=\frac{5}{2}$~~
$Y_4=\frac{4}{3} $}  \\  \hline
 $D_1D_2J_4I_4$ & 23315 \\  \hline
 6~ 6~ 1~ 5 & $-1$  \\ \hline
\end{tabular}

\begin{tabular}{c|cccccccccccccc}  \hline
\multicolumn{ 7}{c}{$Y= 2$~~  $I=0          $~~  $J=\frac{3}{2}$~~
$Y_4=\frac{4}{3} $}  \\  \hline
 $D_1D_2J_4I_4$ & 12325 & 22325 & 24325 & 32325 & 34325 & 23315 \\  \hline
 1~ 6~ 2~ 5 & $-\frac{  1}{   6}$ & $ 0$ & $ 0$ & $-\frac{  1}{  12}$ & $-\frac{  3}{   4}$ & $ 0$  \\
 4~ 9~ 2~ 5 & $ \frac{  1}{   2}$ & $ 0$ & $ 0$ & $ \frac{  1}{   4}$ & $-\frac{  1}{   4}$ & $ 0$  \\
 6~ 1~ 2~ 5 & $ 0$ & $ \frac{  1}{   4}$ & $-\frac{  3}{   4}$ & $ 0$ & $ 0$ & $ 0$  \\
 6~ 6~ 1~ 5 & $ 0$ & $ 0$ & $ 0$ & $ 0$ & $ 0$ & $-1$  \\
 6~ 6~ 2~ 5 & $-\frac{  1}{   3}$ & $ 0$ & $ 0$ & $ \frac{  2}{   3}$ & $ 0$ & $ 0$  \\
 9~ 4~ 2~ 5 & $ 0$ & $-\frac{  3}{   4}$ & $-\frac{  1}{   4}$ & $ 0$ & $ 0$ & $ 0$  \\ \hline
\end{tabular}

\begin{tabular}{c|cccccccccccccc}  \hline
\multicolumn{ 8}{c}{$Y= 2$~~  $I=0          $~~  $J=\frac{1}{2}$~~
$Y_4=\frac{4}{3} $}  \\  \hline
 $D_1D_2J_4I_4$ & 12325 & 22325 & 24325 & 32325 & 34325 & 21335 & 23335 \\  \hline
 1~ 6~ 2~ 5 & $-\frac{  1}{   6}$ & $ 0$ & $ 0$ & $-\frac{  1}{  12}$ & $-\frac{  3}{   4}$ & $ 0$ & $ 0$  \\
 4~ 9~ 2~ 5 & $ \frac{  1}{   2}$ & $ 0$ & $ 0$ & $ \frac{  1}{   4}$ & $-\frac{  1}{   4}$ & $ 0$ & $ 0$  \\
 6~ 1~ 2~ 5 & $ 0$ & $ \frac{  1}{   4}$ & $-\frac{  3}{   4}$ & $ 0$ & $ 0$ & $ 0$ & $ 0$  \\
 6~ 6~ 2~ 5 & $-\frac{  1}{   3}$ & $ 0$ & $ 0$ & $ \frac{  2}{   3}$ & $ 0$ & $ 0$ & $ 0$  \\
 6~ 6~ 3~ 5 & $ 0$ & $ 0$ & $ 0$ & $ 0$ & $ 0$ & $ \frac{  1}{   2}$ & $-\frac{  1}{   2}$  \\
 9~ 4~ 2~ 5 & $ 0$ & $-\frac{  3}{   4}$ & $-\frac{  1}{   4}$ & $ 0$ & $ 0$ & $ 0$ & $ 0$  \\
 9~ 9~ 3~ 5 & $ 0$ & $ 0$ & $ 0$ & $ 0$ & $ 0$ & $ \frac{  1}{   2}$ & $ \frac{  1}{   2}$  \\ \hline
\end{tabular}

\begin{tabular}{c|cccccccccccccc}  \hline
\multicolumn{ 4}{c}{$Y= 1$~~  $I=\frac{5}{2}$~~  $J=\frac{5}{2}$~~
$Y_4=\frac{4}{3} $}  \\  \hline
 $D_1D_2J_4I_4$ & 21111 \\  \hline
 6~ 6~ 1~ 1 & $ 1$  \\ \hline
\end{tabular}

\begin{tabular}{c|cccccccccccccc}  \hline
\multicolumn{ 5}{c}{$Y= 1$~~  $I=\frac{5}{2}$~~  $J=\frac{3}{2}$~~
$Y_4=\frac{4}{3} $}  \\  \hline
 $D_1D_2J_4I_4$ & 12121 & 22121 & 32121 & 21111 \\  \hline
 1~ 6~ 2~ 1 & $ \frac{  2}{   3}$ & $ 0$ & $ \frac{  1}{   3}$ & $ 0$  \\
 6~ 1~ 2~ 1 & $ 0$ & $-1$ & $ 0$ & $ 0$  \\
 6~ 6~ 1~ 1 & $ 0$ & $ 0$ & $ 0$ & $ 1$  \\
 6~ 6~ 2~ 1 & $-\frac{  1}{   3}$ & $ 0$ & $ \frac{  2}{   3}$ & $ 0$  \\ \hline
\end{tabular}

\begin{tabular}{c|cccccccccccccc}  \hline
\multicolumn{ 5}{c}{$Y= 1$~~  $I=\frac{5}{2}$~~  $J=\frac{1}{2}$~~
$Y_4=\frac{4}{3} $}  \\  \hline
 $D_1D_2J_4I_4$ & 12121 & 22121 & 32121 & 23131 \\  \hline
 1~ 6~ 2~ 1 & $ \frac{  2}{   3}$ & $ 0$ & $ \frac{  1}{   3}$ & $ 0$  \\
 6~ 1~ 2~ 1 & $ 0$ & $-1$ & $ 0$ & $ 0$  \\
 6~ 6~ 2~ 1 & $-\frac{  1}{   3}$ & $ 0$ & $ \frac{  2}{   3}$ & $ 0$  \\
 6~ 6~ 3~ 1 & $ 0$ & $ 0$ & $ 0$ & $-1$  \\ \hline
\end{tabular}

\begin{tabular}{c|cccccccccccccc}  \hline
\multicolumn{ 5}{c}{$Y= 1$~~  $I=\frac{3}{2}$~~  $J=\frac{5}{2}$~~
$Y_4=\frac{4}{3} $}  \\  \hline
 $D_1D_2J_4I_4$ & 12213 & 22213 & 32213 & 21111 \\  \hline
 4~ 6~ 1~ 3 & $ \frac{  2}{   3}$ & $ 0$ & $ \frac{  1}{   3}$ & $ 0$  \\
 6~ 4~ 1~ 3 & $ 0$ & $-1$ & $ 0$ & $ 0$  \\
 6~ 6~ 1~ 1 & $ 0$ & $ 0$ & $ 0$ & $ 1$  \\
 6~ 6~ 1~ 3 & $-\frac{  1}{   3}$ & $ 0$ & $ \frac{  2}{   3}$ & $ 0$  \\ \hline
\end{tabular}

\begin{tabular}{c|cccccccccccccc}  \hline
\multicolumn{ 5}{c}{$Y= 1$~~  $I=\frac{3}{2}$~~  $J=\frac{5}{2}$~~
$Y_4=\frac{1}{3} $}  \\  \hline
 $D_1D_2J_4I_4$ & 12212 & 22212 & 32212 & 21112 \\  \hline
 5~ 6~ 1~ 2 & $ \frac{  2}{   3}$ & $ 0$ & $ \frac{  1}{   3}$ & $ 0$  \\
 6~ 5~ 1~ 2 & $ 0$ & $-1$ & $ 0$ & $ 0$  \\
 6~ 7~ 1~ 2 & $ 0$ & $ 0$ & $ 0$ & $ 1$  \\
 7~ 6~ 1~ 2 & $-\frac{  1}{   3}$ & $ 0$ & $ \frac{  2}{   3}$ & $ 0$  \\ \hline
\end{tabular}

\begin{tabular}{c|cccccccccccccc}  \hline
\multicolumn{15}{c}{$Y= 1$~~  $I=\frac{3}{2}$~~  $J=\frac{3}{2}$~~
$Y_4=\frac{4}{3} $}  \\  \hline
 $D_1D_2J_4I_4$ & 12121 & 22121 & 32121 & 12223 & 22223 & 24223 & 32223 & 34223 & 12213 & 22213 & 32213 & 21111 & 21223 & 23223 \\  \hline
 1~ 6~ 2~ 1 & $ \frac{  2}{   3}$ & $ 0$ & $ \frac{  1}{   3}$ & $ 0$ & $ 0$ & $ 0$ & $ 0$ & $ 0$ & $ 0$ & $ 0$ & $ 0$ & $ 0$ & $ 0$ & $ 0$  \\
 1~ 6~ 2~ 3 & $ 0$ & $ 0$ & $ 0$ & $ 0$ & $ \frac{  1}{   2}$ & $-\frac{  1}{   2}$ & $ 0$ & $ 0$ & $ 0$ & $ 0$ & $ 0$ & $ 0$ & $ 0$ & $ 0$  \\
 4~ 6~ 1~ 3 & $ 0$ & $ 0$ & $ 0$ & $ 0$ & $ 0$ & $ 0$ & $ 0$ & $ 0$ & $ \frac{  2}{   3}$ & $ 0$ & $ \frac{  1}{   3}$ & $ 0$ & $ 0$ & $ 0$  \\
 4~ 6~ 2~ 3 & $ 0$ & $ 0$ & $ 0$ & $ 0$ & $ \frac{  1}{   2}$ & $ \frac{  1}{   2}$ & $ 0$ & $ 0$ & $ 0$ & $ 0$ & $ 0$ & $ 0$ & $ 0$ & $ 0$  \\
 6~ 1~ 2~ 1 & $ 0$ & $-1$ & $ 0$ & $ 0$ & $ 0$ & $ 0$ & $ 0$ & $ 0$ & $ 0$ & $ 0$ & $ 0$ & $ 0$ & $ 0$ & $ 0$  \\
 6~ 1~ 2~ 3 & $ 0$ & $ 0$ & $ 0$ & $-\frac{  1}{   3}$ & $ 0$ & $ 0$ & $-\frac{  1}{   6}$ & $-\frac{  1}{   2}$ & $ 0$ & $ 0$ & $ 0$ & $ 0$ & $ 0$ & $ 0$  \\
\hline
\end{tabular}

\begin{tabular}{c|cccccccccccccc}  \hline
\multicolumn{15}{c}{$Y= 1$~~  $I=\frac{3}{2}$~~  $J=\frac{3}{2}$~~
$Y_4=\frac{1}{3} $}  \\  \hline
 $D_1D_2J_4I_4$ & 12122 & 12222 & 22122 & 22222 & 24222 & 32122 & 32222 & 34222 & 12212 & 22212 & 32212 & 21112 & 21222 & 23222 \\  \hline
 1~ 7~ 2~ 2 & $ \frac{  1}{   3}$ & $-\frac{  1}{   6}$ & $ 0$ & $ 0$ & $ 0$ & $ \frac{  1}{   6}$ & $-\frac{  1}{  12}$ & $-\frac{  1}{   4}$ & $ 0$ & $ 0$ & $ 0$ & $ 0$ & $ 0$ & $ 0$  \\
 2~ 6~ 2~ 2 & $ \frac{  1}{   3}$ & $ \frac{  1}{   6}$ & $ 0$ & $ 0$ & $ 0$ & $ \frac{  1}{   6}$ & $ \frac{  1}{  12}$ & $ \frac{  1}{   4}$ & $ 0$ & $ 0$ & $ 0$ & $ 0$ & $ 0$ & $ 0$  \\
 5~ 6~ 1~ 2 & $ 0$ & $ 0$ & $ 0$ & $ 0$ & $ 0$ & $ 0$ & $ 0$ & $ 0$ & $ \frac{  2}{   3}$ & $ 0$ & $ \frac{  1}{   3}$ & $ 0$ & $ 0$ & $ 0$  \\
 5~ 6~ 2~ 2 & $ 0$ & $ 0$ & $ 0$ & $ \frac{  1}{   2}$ & $ \frac{  1}{   2}$ & $ 0$ & $ 0$ & $ 0$ & $ 0$ & $ 0$ & $ 0$ & $ 0$ & $ 0$ & $ 0$  \\
 6~ 2~ 2~ 2 & $ 0$ & $ 0$ & $-\frac{  1}{   2}$ & $-\frac{  1}{   4}$ & $ \frac{  1}{   4}$ & $ 0$ & $ 0$ & $ 0$ & $ 0$ & $ 0$ & $ 0$ & $ 0$ & $ 0$ & $ 0$  \\
 6~ 5~ 1~ 2 & $ 0$ & $ 0$ & $ 0$ & $ 0$ & $ 0$ & $ 0$ & $ 0$ & $ 0$ & $ 0$ & $-1$ & $ 0$ & $ 0$ & $ 0$ & $ 0$  \\
\hline
\end{tabular}

\begin{tabular}{c|cccccccccccccc}  \hline
\multicolumn{15}{c}{$Y= 1$~~  $I=\frac{3}{2}$~~  $J=\frac{1}{2}$~~
$Y_4=\frac{4}{3} $}  \\  \hline
 $D_1D_2J_4I_4$ & 12121 & 22121 & 32121 & 12223 & 22223 & 24223 & 32223 & 34223 & 12233 & 22233 & 24233 & 32233 & 34233 & 21223 \\  \hline
 1~ 6~ 2~ 1 & $ \frac{  2}{   3}$ & $ 0$ & $ \frac{  1}{   3}$ & $ 0$ & $ 0$ & $ 0$ & $ 0$ & $ 0$ & $ 0$ & $ 0$ & $ 0$ & $ 0$ & $ 0$ & $ 0$  \\
 1~ 6~ 2~ 3 & $ 0$ & $ 0$ & $ 0$ & $ 0$ & $ \frac{  1}{   2}$ & $-\frac{  1}{   2}$ & $ 0$ & $ 0$ & $ 0$ & $ 0$ & $ 0$ & $ 0$ & $ 0$ & $ 0$  \\
 1~ 9~ 3~ 3 & $ 0$ & $ 0$ & $ 0$ & $ 0$ & $ 0$ & $ 0$ & $ 0$ & $ 0$ & $ \frac{  1}{   2}$ & $ 0$ & $ 0$ & $ \frac{  1}{   4}$ & $-\frac{  1}{   4}$ & $ 0$  \\
 4~ 6~ 2~ 3 & $ 0$ & $ 0$ & $ 0$ & $ 0$ & $ \frac{  1}{   2}$ & $ \frac{  1}{   2}$ & $ 0$ & $ 0$ & $ 0$ & $ 0$ & $ 0$ & $ 0$ & $ 0$ & $ 0$  \\
 4~ 6~ 3~ 3 & $ 0$ & $ 0$ & $ 0$ & $ 0$ & $ 0$ & $ 0$ & $ 0$ & $ 0$ & $-\frac{  1}{   6}$ & $ 0$ & $ 0$ & $-\frac{  1}{  12}$ & $-\frac{  3}{   4}$ & $ 0$  \\
 6~ 1~ 2~ 1 & $ 0$ & $-1$ & $ 0$ & $ 0$ & $ 0$ & $ 0$ & $ 0$ & $ 0$ & $ 0$ & $ 0$ & $ 0$ & $ 0$ & $ 0$ & $ 0$  \\
\hline
\end{tabular}

\begin{tabular}{c|cccccccccccccc}  \hline
 $D_1D_2J_4I_4$ & 23223 & 23131 \\  \hline
 1~ 6~ 2~ 1 & $ 0$ & $ 0$  \\
 1~ 6~ 2~ 3 & $ 0$ & $ 0$  \\
 1~ 9~ 3~ 3 & $ 0$ & $ 0$  \\
 4~ 6~ 2~ 3 & $ 0$ & $ 0$  \\
 4~ 6~ 3~ 3 & $ 0$ & $ 0$  \\
 6~ 1~ 2~ 1 & $ 0$ & $ 0$  \\
 6~ 1~ 2~ 3 & $ 0$ & $ 0$  \\
 6~ 4~ 2~ 3 & $ 0$ & $ 0$  \\
 6~ 4~ 3~ 3 & $ 0$ & $ 0$  \\
 6~ 6~ 2~ 1 & $ 0$ & $ 0$  \\
 6~ 6~ 2~ 3 & $-\frac{  2}{   3}$ & $ 0$  \\
 6~ 6~ 3~ 1 & $ 0$ & $-1$  \\
 6~ 6~ 3~ 3 & $ 0$ & $ 0$  \\
 6~ 9~ 2~ 3 & $ \frac{  1}{   3}$ & $ 0$  \\
 9~ 1~ 3~ 3 & $ 0$ & $ 0$  \\
 9~ 6~ 2~ 3 & $ 0$ & $ 0$  \\ \hline
\end{tabular}

\begin{tabular}{c|cccccccccccccc}  \hline
\multicolumn{15}{c}{$Y= 1$~~  $I=\frac{3}{2}$~~  $J=\frac{1}{2}$~~
$Y_4=\frac{1}{3} $}  \\  \hline
 $D_1D_2J_4I_4$ & 12122 & 12222 & 22122 & 22222 & 24222 & 32122 & 32222 & 34222 & 12232 & 22232 & 24232 & 32232 & 34232 & 21222 \\  \hline
 1~ 7~ 2~ 2 & $ \frac{  1}{   3}$ & $-\frac{  1}{   6}$ & $ 0$ & $ 0$ & $ 0$ & $ \frac{  1}{   6}$ & $-\frac{  1}{  12}$ & $-\frac{  1}{   4}$ & $ 0$ & $ 0$ & $ 0$ & $ 0$ & $ 0$ & $ 0$  \\
 1~10~ 3~ 2 & $ 0$ & $ 0$ & $ 0$ & $ 0$ & $ 0$ & $ 0$ & $ 0$ & $ 0$ & $ \frac{  1}{   2}$ & $ 0$ & $ 0$ & $ \frac{  1}{   4}$ & $-\frac{  1}{   4}$ & $ 0$  \\
 2~ 6~ 2~ 2 & $ \frac{  1}{   3}$ & $ \frac{  1}{   6}$ & $ 0$ & $ 0$ & $ 0$ & $ \frac{  1}{   6}$ & $ \frac{  1}{  12}$ & $ \frac{  1}{   4}$ & $ 0$ & $ 0$ & $ 0$ & $ 0$ & $ 0$ & $ 0$  \\
 5~ 6~ 2~ 2 & $ 0$ & $ 0$ & $ 0$ & $ \frac{  1}{   2}$ & $ \frac{  1}{   2}$ & $ 0$ & $ 0$ & $ 0$ & $ 0$ & $ 0$ & $ 0$ & $ 0$ & $ 0$ & $ 0$  \\
 5~ 6~ 3~ 2 & $ 0$ & $ 0$ & $ 0$ & $ 0$ & $ 0$ & $ 0$ & $ 0$ & $ 0$ & $-\frac{  1}{   6}$ & $ 0$ & $ 0$ & $-\frac{  1}{  12}$ & $-\frac{  3}{   4}$ & $ 0$  \\
 6~ 2~ 2~ 2 & $ 0$ & $ 0$ & $-\frac{  1}{   2}$ & $-\frac{  1}{   4}$ & $ \frac{  1}{   4}$ & $ 0$ & $ 0$ & $ 0$ & $ 0$ & $ 0$ & $ 0$ & $ 0$ & $ 0$ & $ 0$  \\
\hline
\end{tabular}

\begin{tabular}{c|cccccccccccccc}  \hline
 $D_1D_2J_4I_4$ & 23222 & 23132 \\  \hline
 1~ 7~ 2~ 2 & $ 0$ & $ 0$  \\
 1~10~ 3~ 2 & $ 0$ & $ 0$  \\
 2~ 6~ 2~ 2 & $ 0$ & $ 0$  \\
 5~ 6~ 2~ 2 & $ 0$ & $ 0$  \\
 5~ 6~ 3~ 2 & $ 0$ & $ 0$  \\
 6~ 2~ 2~ 2 & $ 0$ & $ 0$  \\
 6~ 5~ 2~ 2 & $ 0$ & $ 0$  \\
 6~ 5~ 3~ 2 & $ 0$ & $ 0$  \\
 6~ 7~ 2~ 2 & $ 0$ & $ 0$  \\
 6~ 7~ 3~ 2 & $ 0$ & $-1$  \\
 6~10~ 2~ 2 & $ \frac{  1}{   3}$ & $ 0$  \\
 7~ 1~ 2~ 2 & $ 0$ & $ 0$  \\
 7~ 6~ 2~ 2 & $-\frac{  2}{   3}$ & $ 0$  \\
 7~ 6~ 3~ 2 & $ 0$ & $ 0$  \\
10~ 1~ 3~ 2 & $ 0$ & $ 0$  \\
10~ 6~ 2~ 2 & $ 0$ & $ 0$  \\ \hline
\end{tabular}

\begin{tabular}{c|cccccccccccccc}  \hline
\multicolumn{ 5}{c}{$Y= 1$~~  $I=\frac{1}{2}$~~  $J=\frac{5}{2}$~~
$Y_4=\frac{4}{3} $}  \\  \hline
 $D_1D_2J_4I_4$ & 12213 & 22213 & 32213 & 23315 \\  \hline
 4~ 6~ 1~ 3 & $ \frac{  2}{   3}$ & $ 0$ & $ \frac{  1}{   3}$ & $ 0$  \\
 6~ 4~ 1~ 3 & $ 0$ & $-1$ & $ 0$ & $ 0$  \\
 6~ 6~ 1~ 3 & $-\frac{  1}{   3}$ & $ 0$ & $ \frac{  2}{   3}$ & $ 0$  \\
 6~ 6~ 1~ 5 & $ 0$ & $ 0$ & $ 0$ & $-1$  \\ \hline
\end{tabular}

\begin{tabular}{c|cccccccccccccc}  \hline
\multicolumn{ 7}{c}{$Y= 1$~~  $I=\frac{1}{2}$~~  $J=\frac{5}{2}$~~
$Y_4=\frac{1}{3} $}  \\  \hline
 $D_1D_2J_4I_4$ & 12214 & 22214 & 24414 & 32214 & 34414 & 23314 \\  \hline
 4~ 7~ 1~ 4 & $ \frac{  1}{   2}$ & $ 0$ & $ 0$ & $ \frac{  1}{   4}$ & $-\frac{  1}{   4}$ & $ 0$  \\
 5~ 6~ 1~ 4 & $ 0$ & $-\frac{  1}{   4}$ & $ \frac{  3}{   4}$ & $ 0$ & $ 0$ & $ 0$  \\
 6~ 5~ 1~ 4 & $ \frac{  1}{   6}$ & $ 0$ & $ 0$ & $ \frac{  1}{  12}$ & $ \frac{  3}{   4}$ & $ 0$  \\
 6~ 7~ 1~ 4 & $-\frac{  1}{   3}$ & $ 0$ & $ 0$ & $ \frac{  2}{   3}$ & $ 0$ & $ 0$  \\
 7~ 4~ 1~ 4 & $ 0$ & $-\frac{  3}{   4}$ & $-\frac{  1}{   4}$ & $ 0$ & $ 0$ & $ 0$  \\
 7~ 6~ 1~ 4 & $ 0$ & $ 0$ & $ 0$ & $ 0$ & $ 0$ & $-1$  \\ \hline
\end{tabular}

\begin{tabular}{c|cccccccccccccc}  \hline
\multicolumn{15}{c}{$Y= 1$~~  $I=\frac{1}{2}$~~  $J=\frac{3}{2}$~~
$Y_4=\frac{4}{3} $}  \\  \hline
 $D_1D_2J_4I_4$ & 12223 & 22223 & 24223 & 32223 & 34223 & 12325 & 22325 & 24325 & 32325 & 34325 & 12213 & 22213 & 32213 & 23315 \\  \hline
 1~ 6~ 2~ 3 & $ 0$ & $ \frac{  1}{   2}$ & $-\frac{  1}{   2}$ & $ 0$ & $ 0$ & $ 0$ & $ 0$ & $ 0$ & $ 0$ & $ 0$ & $ 0$ & $ 0$ & $ 0$ & $ 0$  \\
 1~ 6~ 2~ 5 & $ 0$ & $ 0$ & $ 0$ & $ 0$ & $ 0$ & $-\frac{  1}{   6}$ & $ 0$ & $ 0$ & $-\frac{  1}{  12}$ & $-\frac{  3}{   4}$ & $ 0$ & $ 0$ & $ 0$ & $ 0$  \\
 4~ 6~ 1~ 3 & $ 0$ & $ 0$ & $ 0$ & $ 0$ & $ 0$ & $ 0$ & $ 0$ & $ 0$ & $ 0$ & $ 0$ & $ \frac{  2}{   3}$ & $ 0$ & $ \frac{  1}{   3}$ & $ 0$  \\
 4~ 6~ 2~ 3 & $ 0$ & $ \frac{  1}{   2}$ & $ \frac{  1}{   2}$ & $ 0$ & $ 0$ & $ 0$ & $ 0$ & $ 0$ & $ 0$ & $ 0$ & $ 0$ & $ 0$ & $ 0$ & $ 0$  \\
 4~ 9~ 2~ 5 & $ 0$ & $ 0$ & $ 0$ & $ 0$ & $ 0$ & $ \frac{  1}{   2}$ & $ 0$ & $ 0$ & $ \frac{  1}{   4}$ & $-\frac{  1}{   4}$ & $ 0$ & $ 0$ & $ 0$ & $ 0$  \\
 6~ 1~ 2~ 3 & $-\frac{  1}{   3}$ & $ 0$ & $ 0$ & $-\frac{  1}{   6}$ & $-\frac{  1}{   2}$ & $ 0$ & $ 0$ & $ 0$ & $ 0$ & $ 0$ & $ 0$ & $ 0$ & $ 0$ & $ 0$  \\
\hline
\end{tabular}

\begin{tabular}{c|cccccccccccccc}  \hline
 $D_1D_2J_4I_4$ & 21223 & 23223 \\  \hline
 1~ 6~ 2~ 3 & $ 0$ & $ 0$  \\
 1~ 6~ 2~ 5 & $ 0$ & $ 0$  \\
 4~ 6~ 1~ 3 & $ 0$ & $ 0$  \\
 4~ 6~ 2~ 3 & $ 0$ & $ 0$  \\
 4~ 9~ 2~ 5 & $ 0$ & $ 0$  \\
 6~ 1~ 2~ 3 & $ 0$ & $ 0$  \\
 6~ 1~ 2~ 5 & $ 0$ & $ 0$  \\
 6~ 4~ 1~ 3 & $ 0$ & $ 0$  \\
 6~ 4~ 2~ 3 & $ 0$ & $ 0$  \\
 6~ 6~ 1~ 3 & $ 0$ & $ 0$  \\
 6~ 6~ 1~ 5 & $ 0$ & $ 0$  \\
 6~ 6~ 2~ 3 & $ \frac{  1}{   3}$ & $-\frac{  2}{   3}$  \\
 6~ 6~ 2~ 5 & $ 0$ & $ 0$  \\
 6~ 9~ 2~ 3 & $ \frac{  2}{   3}$ & $ \frac{  1}{   3}$  \\
 9~ 4~ 2~ 5 & $ 0$ & $ 0$  \\
 9~ 6~ 2~ 3 & $ 0$ & $ 0$  \\ \hline
\end{tabular}

\begin{tabular}{c|cccccccccccccc}  \hline
\multicolumn{15}{c}{$Y= 1$~~  $I=\frac{1}{2}$~~  $J=\frac{3}{2}$~~
$Y_4=\frac{1}{3} $}  \\  \hline
 $D_1D_2J_4I_4$ & 12224 & 12324 & 22224 & 22324 & 24224 & 24324 & 32224 & 32324 & 34224 & 34324 & 12214 & 22214 & 24414 & 32214 \\  \hline
 1~ 7~ 2~ 4 & $ 0$ & $ 0$ & $ \frac{  1}{   4}$ & $ \frac{  1}{   8}$ & $-\frac{  1}{   4}$ & $-\frac{  3}{   8}$ & $ 0$ & $ 0$ & $ 0$ & $ 0$ & $ 0$ & $ 0$ & $ 0$ & $ 0$  \\
 2~ 6~ 2~ 4 & $ 0$ & $ 0$ & $ \frac{  1}{   4}$ & $-\frac{  1}{   8}$ & $-\frac{  1}{   4}$ & $ \frac{  3}{   8}$ & $ 0$ & $ 0$ & $ 0$ & $ 0$ & $ 0$ & $ 0$ & $ 0$ & $ 0$  \\
 4~ 7~ 1~ 4 & $ 0$ & $ 0$ & $ 0$ & $ 0$ & $ 0$ & $ 0$ & $ 0$ & $ 0$ & $ 0$ & $ 0$ & $ \frac{  1}{   2}$ & $ 0$ & $ 0$ & $ \frac{  1}{   4}$  \\
 4~ 7~ 2~ 4 & $ 0$ & $ 0$ & $ \frac{  3}{   8}$ & $ 0$ & $ \frac{  3}{   8}$ & $ 0$ & $ 0$ & $ 0$ & $ 0$ & $ 0$ & $ 0$ & $ 0$ & $ 0$ & $ 0$  \\
 4~10~ 2~ 4 & $ 0$ & $ \frac{  1}{   4}$ & $ 0$ & $ 0$ & $ 0$ & $ 0$ & $ 0$ & $ \frac{  1}{   8}$ & $ 0$ & $-\frac{  1}{   8}$ & $ 0$ & $ 0$ & $ 0$ & $ 0$  \\
 5~ 6~ 1~ 4 & $ 0$ & $ 0$ & $ 0$ & $ 0$ & $ 0$ & $ 0$ & $ 0$ & $ 0$ & $ 0$ & $ 0$ & $ 0$ & $-\frac{  1}{   4}$ & $ \frac{  3}{   4}$ & $ 0$  \\
\hline
\end{tabular}

\begin{tabular}{c|cccccccccccccc}  \hline
 $D_1D_2J_4I_4$ & 34414 & 12424 & 22424 & 24424 & 32424 & 34424 & 23314 & 21224 & 23224 & 23424 \\  \hline
 1~ 7~ 2~ 4 & $ 0$ & $ 0$ & $ 0$ & $ 0$ & $ 0$ & $ 0$ & $ 0$ & $ 0$ & $ 0$ & $ 0$  \\
 2~ 6~ 2~ 4 & $ 0$ & $ 0$ & $ 0$ & $ 0$ & $ 0$ & $ 0$ & $ 0$ & $ 0$ & $ 0$ & $ 0$  \\
 4~ 7~ 1~ 4 & $-\frac{  1}{   4}$ & $ 0$ & $ 0$ & $ 0$ & $ 0$ & $ 0$ & $ 0$ & $ 0$ & $ 0$ & $ 0$  \\
 4~ 7~ 2~ 4 & $ 0$ & $ 0$ & $ \frac{  1}{   8}$ & $ \frac{  1}{   8}$ & $ 0$ & $ 0$ & $ 0$ & $ 0$ & $ 0$ & $ 0$  \\
 4~10~ 2~ 4 & $ 0$ & $-\frac{  1}{   6}$ & $ 0$ & $ 0$ & $-\frac{  1}{  12}$ & $-\frac{  1}{   4}$ & $ 0$ & $ 0$ & $ 0$ & $ 0$  \\
 5~ 6~ 1~ 4 & $ 0$ & $ 0$ & $ 0$ & $ 0$ & $ 0$ & $ 0$ & $ 0$ & $ 0$ & $ 0$ & $ 0$  \\
 5~ 6~ 2~ 4 & $ 0$ & $ \frac{  1}{   4}$ & $ 0$ & $ 0$ & $ \frac{  1}{   8}$ & $-\frac{  3}{   8}$ & $ 0$ & $ 0$ & $ 0$ & $ 0$  \\
\hline
\end{tabular}

\begin{tabular}{c|cccccccccccccc}  \hline
\multicolumn{15}{c}{$Y= 1$~~  $I=\frac{1}{2}$~~  $J=\frac{1}{2}$~~
$Y_4=\frac{4}{3} $}  \\  \hline
 $D_1D_2J_4I_4$ & 12223 & 22223 & 24223 & 32223 & 34223 & 12325 & 22325 & 24325 & 32325 & 34325 & 12233 & 22233 & 24233 & 32233 \\  \hline
 1~ 6~ 2~ 3 & $ 0$ & $ \frac{  1}{   2}$ & $-\frac{  1}{   2}$ & $ 0$ & $ 0$ & $ 0$ & $ 0$ & $ 0$ & $ 0$ & $ 0$ & $ 0$ & $ 0$ & $ 0$ & $ 0$  \\
 1~ 6~ 2~ 5 & $ 0$ & $ 0$ & $ 0$ & $ 0$ & $ 0$ & $-\frac{  1}{   6}$ & $ 0$ & $ 0$ & $-\frac{  1}{  12}$ & $-\frac{  3}{   4}$ & $ 0$ & $ 0$ & $ 0$ & $ 0$  \\
 1~ 9~ 3~ 3 & $ 0$ & $ 0$ & $ 0$ & $ 0$ & $ 0$ & $ 0$ & $ 0$ & $ 0$ & $ 0$ & $ 0$ & $ \frac{  1}{   2}$ & $ 0$ & $ 0$ & $ \frac{  1}{   4}$  \\
 4~ 6~ 2~ 3 & $ 0$ & $ \frac{  1}{   2}$ & $ \frac{  1}{   2}$ & $ 0$ & $ 0$ & $ 0$ & $ 0$ & $ 0$ & $ 0$ & $ 0$ & $ 0$ & $ 0$ & $ 0$ & $ 0$  \\
 4~ 6~ 3~ 3 & $ 0$ & $ 0$ & $ 0$ & $ 0$ & $ 0$ & $ 0$ & $ 0$ & $ 0$ & $ 0$ & $ 0$ & $-\frac{  1}{   6}$ & $ 0$ & $ 0$ & $-\frac{  1}{  12}$  \\
 4~ 9~ 2~ 5 & $ 0$ & $ 0$ & $ 0$ & $ 0$ & $ 0$ & $ \frac{  1}{   2}$ & $ 0$ & $ 0$ & $ \frac{  1}{   4}$ & $-\frac{  1}{   4}$ & $ 0$ & $ 0$ & $ 0$ & $ 0$  \\
\hline
\end{tabular}

\begin{tabular}{c|cccccccccccccc}  \hline
 $D_1D_2J_4I_4$ & 34233 & 21223 & 23223 & 21335 & 23335 \\  \hline
 1~ 6~ 2~ 3 & $ 0$ & $ 0$ & $ 0$ & $ 0$ & $ 0$  \\
 1~ 6~ 2~ 5 & $ 0$ & $ 0$ & $ 0$ & $ 0$ & $ 0$  \\
 1~ 9~ 3~ 3 & $-\frac{  1}{   4}$ & $ 0$ & $ 0$ & $ 0$ & $ 0$  \\
 4~ 6~ 2~ 3 & $ 0$ & $ 0$ & $ 0$ & $ 0$ & $ 0$  \\
 4~ 6~ 3~ 3 & $-\frac{  3}{   4}$ & $ 0$ & $ 0$ & $ 0$ & $ 0$  \\
 4~ 9~ 2~ 5 & $ 0$ & $ 0$ & $ 0$ & $ 0$ & $ 0$  \\
 6~ 1~ 2~ 3 & $ 0$ & $ 0$ & $ 0$ & $ 0$ & $ 0$  \\
 6~ 1~ 2~ 5 & $ 0$ & $ 0$ & $ 0$ & $ 0$ & $ 0$  \\
 6~ 4~ 2~ 3 & $ 0$ & $ 0$ & $ 0$ & $ 0$ & $ 0$  \\
 6~ 4~ 3~ 3 & $ 0$ & $ 0$ & $ 0$ & $ 0$ & $ 0$  \\
 6~ 6~ 2~ 3 & $ 0$ & $ \frac{  1}{   3}$ & $-\frac{  2}{   3}$ & $ 0$ & $ 0$  \\
 6~ 6~ 2~ 5 & $ 0$ & $ 0$ & $ 0$ & $ 0$ & $ 0$  \\
 6~ 6~ 3~ 3 & $ 0$ & $ 0$ & $ 0$ & $ 0$ & $ 0$  \\
 6~ 6~ 3~ 5 & $ 0$ & $ 0$ & $ 0$ & $ \frac{  1}{   2}$ & $-\frac{  1}{   2}$  \\
 6~ 9~ 2~ 3 & $ 0$ & $ \frac{  2}{   3}$ & $ \frac{  1}{   3}$ & $ 0$ & $ 0$  \\
 9~ 1~ 3~ 3 & $ 0$ & $ 0$ & $ 0$ & $ 0$ & $ 0$  \\
 9~ 4~ 2~ 5 & $ 0$ & $ 0$ & $ 0$ & $ 0$ & $ 0$  \\
 9~ 6~ 2~ 3 & $ 0$ & $ 0$ & $ 0$ & $ 0$ & $ 0$  \\
 9~ 9~ 3~ 5 & $ 0$ & $ 0$ & $ 0$ & $ \frac{  1}{   2}$ & $ \frac{  1}{   2}$  \\ \hline
\end{tabular}

\begin{tabular}{c|cccccccccccccc}  \hline
\multicolumn{15}{c}{$Y= 1$~~  $I=\frac{1}{2}$~~  $J=\frac{1}{2}$~~
$Y_4=\frac{1}{3} $}  \\  \hline
 $D_1D_2J_4I_4$ & 12224 & 12324 & 22224 & 22324 & 24224 & 24324 & 32224 & 32324 & 34224 & 34324 & 12234 & 12434 & 22234 & 22434 \\  \hline
 1~ 7~ 2~ 4 & $ 0$ & $ 0$ & $ \frac{  1}{   4}$ & $ \frac{  1}{   8}$ & $-\frac{  1}{   4}$ & $-\frac{  3}{   8}$ & $ 0$ & $ 0$ & $ 0$ & $ 0$ & $ 0$ & $ 0$ & $ 0$ & $ 0$  \\
 1~10~ 3~ 4 & $ 0$ & $ 0$ & $ 0$ & $ 0$ & $ 0$ & $ 0$ & $ 0$ & $ 0$ & $ 0$ & $ 0$ & $ 0$ & $ 0$ & $ \frac{  3}{  16}$ & $-\frac{  3}{  16}$  \\
 2~ 6~ 2~ 4 & $ 0$ & $ 0$ & $ \frac{  1}{   4}$ & $-\frac{  1}{   8}$ & $-\frac{  1}{   4}$ & $ \frac{  3}{   8}$ & $ 0$ & $ 0$ & $ 0$ & $ 0$ & $ 0$ & $ 0$ & $ 0$ & $ 0$  \\
 2~ 9~ 3~ 4 & $ 0$ & $ 0$ & $ 0$ & $ 0$ & $ 0$ & $ 0$ & $ 0$ & $ 0$ & $ 0$ & $ 0$ & $ \frac{  3}{   8}$ & $ \frac{  1}{  24}$ & $ 0$ & $ 0$  \\
 4~ 7~ 2~ 4 & $ 0$ & $ 0$ & $ \frac{  3}{   8}$ & $ 0$ & $ \frac{  3}{   8}$ & $ 0$ & $ 0$ & $ 0$ & $ 0$ & $ 0$ & $ 0$ & $ 0$ & $ 0$ & $ 0$  \\
 4~ 7~ 3~ 4 & $ 0$ & $ 0$ & $ 0$ & $ 0$ & $ 0$ & $ 0$ & $ 0$ & $ 0$ & $ 0$ & $ 0$ & $-\frac{  1}{   8}$ & $ \frac{  1}{   8}$ & $ 0$ & $ 0$  \\
\hline
\end{tabular}

\begin{tabular}{c|cccccccccccccc}  \hline
 $D_1D_2J_4I_4$ & 24234 & 24434 & 32234 & 32434 & 34234 & 34434 & 12424 & 22424 & 24424 & 32424 & 34424 & 21224 & 23224 & 21334 \\  \hline
 1~ 7~ 2~ 4 & $ 0$ & $ 0$ & $ 0$ & $ 0$ & $ 0$ & $ 0$ & $ 0$ & $ 0$ & $ 0$ & $ 0$ & $ 0$ & $ 0$ & $ 0$ & $ 0$  \\
 1~10~ 3~ 4 & $ \frac{  1}{  16}$ & $ \frac{  9}{  16}$ & $ 0$ & $ 0$ & $ 0$ & $ 0$ & $ 0$ & $ 0$ & $ 0$ & $ 0$ & $ 0$ & $ 0$ & $ 0$ & $ 0$  \\
 2~ 6~ 2~ 4 & $ 0$ & $ 0$ & $ 0$ & $ 0$ & $ 0$ & $ 0$ & $ 0$ & $ 0$ & $ 0$ & $ 0$ & $ 0$ & $ 0$ & $ 0$ & $ 0$  \\
 2~ 9~ 3~ 4 & $ 0$ & $ 0$ & $ \frac{  3}{  16}$ & $ \frac{  1}{  48}$ & $-\frac{  3}{  16}$ & $ \frac{  3}{  16}$ & $ 0$ & $ 0$ & $ 0$ & $ 0$ & $ 0$ & $ 0$ & $ 0$ & $ 0$  \\
 4~ 7~ 2~ 4 & $ 0$ & $ 0$ & $ 0$ & $ 0$ & $ 0$ & $ 0$ & $ 0$ & $ \frac{  1}{   8}$ & $ \frac{  1}{   8}$ & $ 0$ & $ 0$ & $ 0$ & $ 0$ & $ 0$  \\
 4~ 7~ 3~ 4 & $ 0$ & $ 0$ & $-\frac{  1}{  16}$ & $ \frac{  1}{  16}$ & $-\frac{  9}{  16}$ & $-\frac{  1}{  16}$ & $ 0$ & $ 0$ & $ 0$ & $ 0$ & $ 0$ & $ 0$ & $ 0$ & $ 0$  \\
 4~10~ 2~ 4 & $ 0$ & $ 0$ & $ 0$ & $ 0$ & $ 0$ & $ 0$ & $-\frac{  1}{   6}$ & $ 0$ & $ 0$ & $-\frac{  1}{  12}$ & $-\frac{  1}{   4}$ & $ 0$ & $ 0$ & $ 0$  \\
\hline
\end{tabular}

\begin{tabular}{c|cccccccccccccc}  \hline
 $D_1D_2J_4I_4$ & 23334 & 23424 \\  \hline
 1~ 7~ 2~ 4 & $ 0$ & $ 0$  \\
 1~10~ 3~ 4 & $ 0$ & $ 0$  \\
 2~ 6~ 2~ 4 & $ 0$ & $ 0$  \\
 2~ 9~ 3~ 4 & $ 0$ & $ 0$  \\
 4~ 7~ 2~ 4 & $ 0$ & $ 0$  \\
 4~ 7~ 3~ 4 & $ 0$ & $ 0$  \\
 4~10~ 2~ 4 & $ 0$ & $ 0$  \\
 5~ 6~ 2~ 4 & $ 0$ & $ 0$  \\
 5~ 6~ 3~ 4 & $ 0$ & $ 0$  \\
 5~ 9~ 2~ 4 & $ 0$ & $ 0$  \\
 6~ 2~ 2~ 4 & $ 0$ & $ 0$  \\
 6~ 5~ 2~ 4 & $ 0$ & $ 0$  \\
 6~ 5~ 3~ 4 & $ 0$ & $ 0$  \\
 6~ 7~ 2~ 4 & $ 0$ & $ 0$  \\
 6~ 7~ 3~ 4 & $ 0$ & $ 0$  \\
 6~10~ 2~ 4 & $ 0$ & $ 0$  \\
 7~ 1~ 2~ 4 & $ 0$ & $ 0$  \\
 7~ 4~ 2~ 4 & $ 0$ & $ 0$  \\
 7~ 4~ 3~ 4 & $ 0$ & $ 0$  \\
 7~ 6~ 2~ 4 & $ 0$ & $ 0$  \\
 7~ 6~ 3~ 4 & $-\frac{  1}{   2}$ & $ 0$  \\
 7~ 9~ 2~ 4 & $ 0$ & $-\frac{  1}{   4}$  \\
 9~ 2~ 3~ 4 & $ 0$ & $ 0$  \\
 9~ 5~ 2~ 4 & $ 0$ & $ 0$  \\
 9~ 7~ 2~ 4 & $ 0$ & $ 0$  \\
 9~10~ 3~ 4 & $ \frac{  1}{   2}$ & $ 0$  \\
10~ 1~ 3~ 4 & $ 0$ & $ 0$  \\
10~ 4~ 2~ 4 & $ 0$ & $ 0$  \\
10~ 6~ 2~ 4 & $ 0$ & $-\frac{  3}{   4}$  \\
10~ 9~ 3~ 4 & $ 0$ & $ 0$  \\ \hline
\end{tabular}

\begin{tabular}{c|cccccccccccccc}  \hline
\multicolumn{ 5}{c}{$Y= 0$~~  $I=2          $~~  $J=\frac{5}{2}$~~
$Y_4=\frac{1}{3} $}  \\  \hline
 $D_1D_2J_4I_4$ & 12212 & 22212 & 32212 & 21112 \\  \hline
 5~ 6~ 1~ 2 & $ \frac{  2}{   3}$ & $ 0$ & $ \frac{  1}{   3}$ & $ 0$  \\
 6~ 5~ 1~ 2 & $ 0$ & $-1$ & $ 0$ & $ 0$  \\
 6~ 7~ 1~ 2 & $ 0$ & $ 0$ & $ 0$ & $ 1$  \\
 7~ 6~ 1~ 2 & $-\frac{  1}{   3}$ & $ 0$ & $ \frac{  2}{   3}$ & $ 0$  \\ \hline
\end{tabular}

\begin{tabular}{c|cccccccccccccc}  \hline
\multicolumn{15}{c}{$Y= 0$~~  $I=2          $~~  $J=\frac{3}{2}$~~
$Y_4=\frac{1}{3} $}  \\  \hline
 $D_1D_2J_4I_4$ & 12122 & 12222 & 22122 & 22222 & 24222 & 32122 & 32222 & 34222 & 12212 & 22212 & 32212 & 21112 & 21222 & 23222 \\  \hline
 1~ 7~ 2~ 2 & $ \frac{  1}{   3}$ & $-\frac{  1}{   6}$ & $ 0$ & $ 0$ & $ 0$ & $ \frac{  1}{   6}$ & $-\frac{  1}{  12}$ & $-\frac{  1}{   4}$ & $ 0$ & $ 0$ & $ 0$ & $ 0$ & $ 0$ & $ 0$  \\
 2~ 6~ 2~ 2 & $ \frac{  1}{   3}$ & $ \frac{  1}{   6}$ & $ 0$ & $ 0$ & $ 0$ & $ \frac{  1}{   6}$ & $ \frac{  1}{  12}$ & $ \frac{  1}{   4}$ & $ 0$ & $ 0$ & $ 0$ & $ 0$ & $ 0$ & $ 0$  \\
 5~ 6~ 1~ 2 & $ 0$ & $ 0$ & $ 0$ & $ 0$ & $ 0$ & $ 0$ & $ 0$ & $ 0$ & $ \frac{  2}{   3}$ & $ 0$ & $ \frac{  1}{   3}$ & $ 0$ & $ 0$ & $ 0$  \\
 5~ 6~ 2~ 2 & $ 0$ & $ 0$ & $ 0$ & $ \frac{  1}{   2}$ & $ \frac{  1}{   2}$ & $ 0$ & $ 0$ & $ 0$ & $ 0$ & $ 0$ & $ 0$ & $ 0$ & $ 0$ & $ 0$  \\
 6~ 2~ 2~ 2 & $ 0$ & $ 0$ & $-\frac{  1}{   2}$ & $-\frac{  1}{   4}$ & $ \frac{  1}{   4}$ & $ 0$ & $ 0$ & $ 0$ & $ 0$ & $ 0$ & $ 0$ & $ 0$ & $ 0$ & $ 0$  \\
 6~ 5~ 1~ 2 & $ 0$ & $ 0$ & $ 0$ & $ 0$ & $ 0$ & $ 0$ & $ 0$ & $ 0$ & $ 0$ & $-1$ & $ 0$ & $ 0$ & $ 0$ & $ 0$  \\
\hline
\end{tabular}

\begin{tabular}{c|cccccccccccccc}  \hline
\multicolumn{15}{c}{$Y= 0$~~  $I=2          $~~  $J=\frac{1}{2}$~~
$Y_4=\frac{1}{3} $}  \\  \hline
 $D_1D_2J_4I_4$ & 12122 & 12222 & 22122 & 22222 & 24222 & 32122 & 32222 & 34222 & 12232 & 22232 & 24232 & 32232 & 34232 & 21222 \\  \hline
 1~ 7~ 2~ 2 & $ \frac{  1}{   3}$ & $-\frac{  1}{   6}$ & $ 0$ & $ 0$ & $ 0$ & $ \frac{  1}{   6}$ & $-\frac{  1}{  12}$ & $-\frac{  1}{   4}$ & $ 0$ & $ 0$ & $ 0$ & $ 0$ & $ 0$ & $ 0$  \\
 1~10~ 3~ 2 & $ 0$ & $ 0$ & $ 0$ & $ 0$ & $ 0$ & $ 0$ & $ 0$ & $ 0$ & $ \frac{  1}{   2}$ & $ 0$ & $ 0$ & $ \frac{  1}{   4}$ & $-\frac{  1}{   4}$ & $ 0$  \\
 2~ 6~ 2~ 2 & $ \frac{  1}{   3}$ & $ \frac{  1}{   6}$ & $ 0$ & $ 0$ & $ 0$ & $ \frac{  1}{   6}$ & $ \frac{  1}{  12}$ & $ \frac{  1}{   4}$ & $ 0$ & $ 0$ & $ 0$ & $ 0$ & $ 0$ & $ 0$  \\
 5~ 6~ 2~ 2 & $ 0$ & $ 0$ & $ 0$ & $ \frac{  1}{   2}$ & $ \frac{  1}{   2}$ & $ 0$ & $ 0$ & $ 0$ & $ 0$ & $ 0$ & $ 0$ & $ 0$ & $ 0$ & $ 0$  \\
 5~ 6~ 3~ 2 & $ 0$ & $ 0$ & $ 0$ & $ 0$ & $ 0$ & $ 0$ & $ 0$ & $ 0$ & $-\frac{  1}{   6}$ & $ 0$ & $ 0$ & $-\frac{  1}{  12}$ & $-\frac{  3}{   4}$ & $ 0$  \\
 6~ 2~ 2~ 2 & $ 0$ & $ 0$ & $-\frac{  1}{   2}$ & $-\frac{  1}{   4}$ & $ \frac{  1}{   4}$ & $ 0$ & $ 0$ & $ 0$ & $ 0$ & $ 0$ & $ 0$ & $ 0$ & $ 0$ & $ 0$  \\
\hline
\end{tabular}

\begin{tabular}{c|cccccccccccccc}  \hline
 $D_1D_2J_4I_4$ & 23222 & 23132 \\  \hline
 1~ 7~ 2~ 2 & $ 0$ & $ 0$  \\
 1~10~ 3~ 2 & $ 0$ & $ 0$  \\
 2~ 6~ 2~ 2 & $ 0$ & $ 0$  \\
 5~ 6~ 2~ 2 & $ 0$ & $ 0$  \\
 5~ 6~ 3~ 2 & $ 0$ & $ 0$  \\
 6~ 2~ 2~ 2 & $ 0$ & $ 0$  \\
 6~ 5~ 2~ 2 & $ 0$ & $ 0$  \\
 6~ 5~ 3~ 2 & $ 0$ & $ 0$  \\
 6~ 7~ 2~ 2 & $ 0$ & $ 0$  \\
 6~ 7~ 3~ 2 & $ 0$ & $-1$  \\
 6~10~ 2~ 2 & $ \frac{  1}{   3}$ & $ 0$  \\
 7~ 1~ 2~ 2 & $ 0$ & $ 0$  \\
 7~ 6~ 2~ 2 & $-\frac{  2}{   3}$ & $ 0$  \\
 7~ 6~ 3~ 2 & $ 0$ & $ 0$  \\
10~ 1~ 3~ 2 & $ 0$ & $ 0$  \\
10~ 6~ 2~ 2 & $ 0$ & $ 0$  \\ \hline
\end{tabular}

\begin{tabular}{c|cccccccccccccc}  \hline
\multicolumn{11}{c}{$Y= 0$~~  $I=1          $~~  $J=\frac{5}{2}$~~
$Y_4=\frac{1}{3} $}  \\  \hline
 $D_1D_2J_4I_4$ & 12214 & 22214 & 24414 & 32214 & 34414 & 12212 & 22212 & 32212 & 21112 & 23314 \\  \hline
 4~ 7~ 1~ 4 & $ \frac{  1}{   2}$ & $ 0$ & $ 0$ & $ \frac{  1}{   4}$ & $-\frac{  1}{   4}$ & $ 0$ & $ 0$ & $ 0$ & $ 0$ & $ 0$  \\
 5~ 6~ 1~ 2 & $ 0$ & $ 0$ & $ 0$ & $ 0$ & $ 0$ & $ \frac{  2}{   3}$ & $ 0$ & $ \frac{  1}{   3}$ & $ 0$ & $ 0$  \\
 5~ 6~ 1~ 4 & $ 0$ & $-\frac{  1}{   4}$ & $ \frac{  3}{   4}$ & $ 0$ & $ 0$ & $ 0$ & $ 0$ & $ 0$ & $ 0$ & $ 0$  \\
 6~ 5~ 1~ 2 & $ 0$ & $ 0$ & $ 0$ & $ 0$ & $ 0$ & $ 0$ & $-1$ & $ 0$ & $ 0$ & $ 0$  \\
 6~ 5~ 1~ 4 & $ \frac{  1}{   6}$ & $ 0$ & $ 0$ & $ \frac{  1}{  12}$ & $ \frac{  3}{   4}$ & $ 0$ & $ 0$ & $ 0$ & $ 0$ & $ 0$  \\
 6~ 7~ 1~ 2 & $ 0$ & $ 0$ & $ 0$ & $ 0$ & $ 0$ & $ 0$ & $ 0$ & $ 0$ & $ 1$ & $ 0$  \\
 6~ 7~ 1~ 4 & $-\frac{  1}{   3}$ & $ 0$ & $ 0$ & $ \frac{  2}{   3}$ & $ 0$ & $ 0$ & $ 0$ & $ 0$ & $ 0$ & $ 0$  \\
 7~ 4~ 1~ 4 & $ 0$ & $-\frac{  3}{   4}$ & $-\frac{  1}{   4}$ & $ 0$ & $ 0$ & $ 0$ & $ 0$ & $ 0$ & $ 0$ & $ 0$  \\
 7~ 6~ 1~ 2 & $ 0$ & $ 0$ & $ 0$ & $ 0$ & $ 0$ & $-\frac{  1}{   3}$ & $ 0$ & $ \frac{  2}{   3}$ & $ 0$ & $ 0$  \\
 7~ 6~ 1~ 4 & $ 0$ & $ 0$ & $ 0$ & $ 0$ & $ 0$ & $ 0$ & $ 0$ & $ 0$ & $ 0$ & $-1$  \\ \hline
\end{tabular}

\begin{tabular}{c|cccccccccccccc}  \hline
\multicolumn{ 6}{c}{$Y= 0$~~  $I=1          $~~  $J=\frac{5}{2}$~~
$Y_4=-\frac{2}{3}$}  \\  \hline
 $D_1D_2J_4I_4$ & 12213 & 22213 & 32213 & 21113 & 23313 \\  \hline
 5~ 7~ 1~ 3 & $ \frac{  2}{   3}$ & $ 0$ & $ \frac{  1}{   3}$ & $ 0$ & $ 0$  \\
 6~ 8~ 1~ 3 & $ 0$ & $ 0$ & $ 0$ & $ \frac{  1}{   3}$ & $-\frac{  2}{   3}$  \\
 7~ 5~ 1~ 3 & $ 0$ & $-1$ & $ 0$ & $ 0$ & $ 0$  \\
 7~ 7~ 1~ 3 & $ 0$ & $ 0$ & $ 0$ & $ \frac{  2}{   3}$ & $ \frac{  1}{   3}$  \\
 8~ 6~ 1~ 3 & $-\frac{  1}{   3}$ & $ 0$ & $ \frac{  2}{   3}$ & $ 0$ & $ 0$  \\ \hline
\end{tabular}

\begin{tabular}{c|cccccccccccccc}  \hline
\multicolumn{15}{c}{$Y= 0$~~  $I=1          $~~  $J=\frac{3}{2}$~~
$Y_4=\frac{1}{3} $}  \\  \hline
 $D_1D_2J_4I_4$ & 12122 & 12222 & 22122 & 22222 & 24222 & 32122 & 32222 & 34222 & 12224 & 12324 & 22224 & 22324 & 24224 & 24324 \\  \hline
 1~ 7~ 2~ 2 & $ \frac{  1}{   3}$ & $-\frac{  1}{   6}$ & $ 0$ & $ 0$ & $ 0$ & $ \frac{  1}{   6}$ & $-\frac{  1}{  12}$ & $-\frac{  1}{   4}$ & $ 0$ & $ 0$ & $ 0$ & $ 0$ & $ 0$ & $ 0$  \\
 1~ 7~ 2~ 4 & $ 0$ & $ 0$ & $ 0$ & $ 0$ & $ 0$ & $ 0$ & $ 0$ & $ 0$ & $ 0$ & $ 0$ & $ \frac{  1}{   4}$ & $ \frac{  1}{   8}$ & $-\frac{  1}{   4}$ & $-\frac{  3}{   8}$  \\
 2~ 6~ 2~ 2 & $ \frac{  1}{   3}$ & $ \frac{  1}{   6}$ & $ 0$ & $ 0$ & $ 0$ & $ \frac{  1}{   6}$ & $ \frac{  1}{  12}$ & $ \frac{  1}{   4}$ & $ 0$ & $ 0$ & $ 0$ & $ 0$ & $ 0$ & $ 0$  \\
 2~ 6~ 2~ 4 & $ 0$ & $ 0$ & $ 0$ & $ 0$ & $ 0$ & $ 0$ & $ 0$ & $ 0$ & $ 0$ & $ 0$ & $ \frac{  1}{   4}$ & $-\frac{  1}{   8}$ & $-\frac{  1}{   4}$ & $ \frac{  3}{   8}$  \\
 4~ 7~ 1~ 4 & $ 0$ & $ 0$ & $ 0$ & $ 0$ & $ 0$ & $ 0$ & $ 0$ & $ 0$ & $ 0$ & $ 0$ & $ 0$ & $ 0$ & $ 0$ & $ 0$  \\
 4~ 7~ 2~ 4 & $ 0$ & $ 0$ & $ 0$ & $ 0$ & $ 0$ & $ 0$ & $ 0$ & $ 0$ & $ 0$ & $ 0$ & $ \frac{  3}{   8}$ & $ 0$ & $ \frac{  3}{   8}$ & $ 0$  \\
\hline
\end{tabular}

\begin{tabular}{c|cccccccccccccc}  \hline
 $D_1D_2J_4I_4$ & 32224 & 32324 & 34224 & 34324 & 12214 & 22214 & 24414 & 32214 & 34414 & 12424 & 22424 & 24424 & 32424 & 34424 \\  \hline
 1~ 7~ 2~ 2 & $ 0$ & $ 0$ & $ 0$ & $ 0$ & $ 0$ & $ 0$ & $ 0$ & $ 0$ & $ 0$ & $ 0$ & $ 0$ & $ 0$ & $ 0$ & $ 0$  \\
 1~ 7~ 2~ 4 & $ 0$ & $ 0$ & $ 0$ & $ 0$ & $ 0$ & $ 0$ & $ 0$ & $ 0$ & $ 0$ & $ 0$ & $ 0$ & $ 0$ & $ 0$ & $ 0$  \\
 2~ 6~ 2~ 2 & $ 0$ & $ 0$ & $ 0$ & $ 0$ & $ 0$ & $ 0$ & $ 0$ & $ 0$ & $ 0$ & $ 0$ & $ 0$ & $ 0$ & $ 0$ & $ 0$  \\
 2~ 6~ 2~ 4 & $ 0$ & $ 0$ & $ 0$ & $ 0$ & $ 0$ & $ 0$ & $ 0$ & $ 0$ & $ 0$ & $ 0$ & $ 0$ & $ 0$ & $ 0$ & $ 0$  \\
 4~ 7~ 1~ 4 & $ 0$ & $ 0$ & $ 0$ & $ 0$ & $ \frac{  1}{   2}$ & $ 0$ & $ 0$ & $ \frac{  1}{   4}$ & $-\frac{  1}{   4}$ & $ 0$ & $ 0$ & $ 0$ & $ 0$ & $ 0$  \\
 4~ 7~ 2~ 4 & $ 0$ & $ 0$ & $ 0$ & $ 0$ & $ 0$ & $ 0$ & $ 0$ & $ 0$ & $ 0$ & $ 0$ & $ \frac{  1}{   8}$ & $ \frac{  1}{   8}$ & $ 0$ & $ 0$  \\
 4~10~ 2~ 4 & $ 0$ & $ \frac{  1}{   8}$ & $ 0$ & $-\frac{  1}{   8}$ & $ 0$ & $ 0$ & $ 0$ & $ 0$ & $ 0$ & $-\frac{  1}{   6}$ & $ 0$ & $ 0$ & $-\frac{  1}{  12}$ & $-\frac{  1}{   4}$  \\
\hline
\end{tabular}

\begin{tabular}{c|cccccccccccccc}  \hline
 $D_1D_2J_4I_4$ & 12212 & 22212 & 32212 & 21112 & 23314 & 21222 & 23222 & 21224 & 23224 & 23424 \\  \hline
 1~ 7~ 2~ 2 & $ 0$ & $ 0$ & $ 0$ & $ 0$ & $ 0$ & $ 0$ & $ 0$ & $ 0$ & $ 0$ & $ 0$  \\
 1~ 7~ 2~ 4 & $ 0$ & $ 0$ & $ 0$ & $ 0$ & $ 0$ & $ 0$ & $ 0$ & $ 0$ & $ 0$ & $ 0$  \\
 2~ 6~ 2~ 2 & $ 0$ & $ 0$ & $ 0$ & $ 0$ & $ 0$ & $ 0$ & $ 0$ & $ 0$ & $ 0$ & $ 0$  \\
 2~ 6~ 2~ 4 & $ 0$ & $ 0$ & $ 0$ & $ 0$ & $ 0$ & $ 0$ & $ 0$ & $ 0$ & $ 0$ & $ 0$  \\
 4~ 7~ 1~ 4 & $ 0$ & $ 0$ & $ 0$ & $ 0$ & $ 0$ & $ 0$ & $ 0$ & $ 0$ & $ 0$ & $ 0$  \\
 4~ 7~ 2~ 4 & $ 0$ & $ 0$ & $ 0$ & $ 0$ & $ 0$ & $ 0$ & $ 0$ & $ 0$ & $ 0$ & $ 0$  \\
 4~10~ 2~ 4 & $ 0$ & $ 0$ & $ 0$ & $ 0$ & $ 0$ & $ 0$ & $ 0$ & $ 0$ & $ 0$ & $ 0$  \\
\hline
\end{tabular}

\begin{tabular}{c|cccccccccccccc}  \hline
\multicolumn{15}{c}{$Y= 0$~~  $I=1          $~~  $J=\frac{3}{2}$~~
$Y_4=-\frac{2}{3}$}  \\  \hline
 $D_1D_2J_4I_4$ & 12123 & 12223 & 12323 & 22123 & 22223 & 22323 & 24223 & 24323 & 32123 & 32223 & 32323 & 34223 & 34323 & 12213 \\  \hline
 1~ 8~ 2~ 3 & $ \frac{  1}{   9}$ & $-\frac{  1}{   6}$ & $-\frac{  1}{  18}$ & $ 0$ & $ 0$ & $ 0$ & $ 0$ & $ 0$ & $ \frac{  1}{  18}$ & $-\frac{  1}{  12}$ & $-\frac{  1}{  36}$ & $-\frac{  1}{   4}$ & $-\frac{  1}{   4}$ & $ 0$  \\
 2~ 7~ 2~ 3 & $ \frac{  4}{   9}$ & $ 0$ & $ \frac{  1}{  18}$ & $ 0$ & $ 0$ & $ 0$ & $ 0$ & $ 0$ & $ \frac{  2}{   9}$ & $ 0$ & $ \frac{  1}{  36}$ & $ 0$ & $ \frac{  1}{   4}$ & $ 0$  \\
 3~ 6~ 2~ 3 & $ \frac{  1}{   9}$ & $ \frac{  1}{   6}$ & $-\frac{  1}{  18}$ & $ 0$ & $ 0$ & $ 0$ & $ 0$ & $ 0$ & $ \frac{  1}{  18}$ & $ \frac{  1}{  12}$ & $-\frac{  1}{  36}$ & $ \frac{  1}{   4}$ & $-\frac{  1}{   4}$ & $ 0$  \\
 5~ 7~ 1~ 3 & $ 0$ & $ 0$ & $ 0$ & $ 0$ & $ 0$ & $ 0$ & $ 0$ & $ 0$ & $ 0$ & $ 0$ & $ 0$ & $ 0$ & $ 0$ & $ \frac{  2}{   3}$  \\
 5~ 7~ 2~ 3 & $ 0$ & $ 0$ & $ 0$ & $ 0$ & $ \frac{  1}{   2}$ & $ 0$ & $ \frac{  1}{   2}$ & $ 0$ & $ 0$ & $ 0$ & $ 0$ & $ 0$ & $ 0$ & $ 0$  \\
 5~10~ 2~ 3 & $ 0$ & $ 0$ & $ \frac{  1}{   2}$ & $ 0$ & $ 0$ & $ 0$ & $ 0$ & $ 0$ & $ 0$ & $ 0$ & $ \frac{  1}{   4}$ & $ 0$ & $-\frac{  1}{   4}$ & $ 0$  \\
\hline
\end{tabular}

\begin{tabular}{c|cccccccccccccc}  \hline
 $D_1D_2J_4I_4$ & 22213 & 32213 & 21113 & 23313 & 21223 & 23223 \\  \hline
 1~ 8~ 2~ 3 & $ 0$ & $ 0$ & $ 0$ & $ 0$ & $ 0$ & $ 0$  \\
 2~ 7~ 2~ 3 & $ 0$ & $ 0$ & $ 0$ & $ 0$ & $ 0$ & $ 0$  \\
 3~ 6~ 2~ 3 & $ 0$ & $ 0$ & $ 0$ & $ 0$ & $ 0$ & $ 0$  \\
 5~ 7~ 1~ 3 & $ 0$ & $ \frac{  1}{   3}$ & $ 0$ & $ 0$ & $ 0$ & $ 0$  \\
 5~ 7~ 2~ 3 & $ 0$ & $ 0$ & $ 0$ & $ 0$ & $ 0$ & $ 0$  \\
 5~10~ 2~ 3 & $ 0$ & $ 0$ & $ 0$ & $ 0$ & $ 0$ & $ 0$  \\
 6~ 3~ 2~ 3 & $ 0$ & $ 0$ & $ 0$ & $ 0$ & $ 0$ & $ 0$  \\
 6~ 8~ 1~ 3 & $ 0$ & $ 0$ & $ \frac{  1}{   3}$ & $-\frac{  2}{   3}$ & $ 0$ & $ 0$  \\
 6~ 8~ 2~ 3 & $ 0$ & $ 0$ & $ 0$ & $ 0$ & $ 0$ & $ 0$  \\
 7~ 2~ 2~ 3 & $ 0$ & $ 0$ & $ 0$ & $ 0$ & $ 0$ & $ 0$  \\
 7~ 5~ 1~ 3 & $-1$ & $ 0$ & $ 0$ & $ 0$ & $ 0$ & $ 0$  \\
 7~ 5~ 2~ 3 & $ 0$ & $ 0$ & $ 0$ & $ 0$ & $ 0$ & $ 0$  \\
 7~ 7~ 1~ 3 & $ 0$ & $ 0$ & $ \frac{  2}{   3}$ & $ \frac{  1}{   3}$ & $ 0$ & $ 0$  \\
 7~ 7~ 2~ 3 & $ 0$ & $ 0$ & $ 0$ & $ 0$ & $ 0$ & $ 0$  \\
 7~10~ 2~ 3 & $ 0$ & $ 0$ & $ 0$ & $ 0$ & $ \frac{  2}{   3}$ & $ \frac{  1}{   3}$  \\
 8~ 1~ 2~ 3 & $ 0$ & $ 0$ & $ 0$ & $ 0$ & $ 0$ & $ 0$  \\
 8~ 6~ 1~ 3 & $ 0$ & $ \frac{  2}{   3}$ & $ 0$ & $ 0$ & $ 0$ & $ 0$  \\
 8~ 6~ 2~ 3 & $ 0$ & $ 0$ & $ 0$ & $ 0$ & $ \frac{  1}{   3}$ & $-\frac{  2}{   3}$  \\
10~ 5~ 2~ 3 & $ 0$ & $ 0$ & $ 0$ & $ 0$ & $ 0$ & $ 0$  \\
10~ 7~ 2~ 3 & $ 0$ & $ 0$ & $ 0$ & $ 0$ & $ 0$ & $ 0$  \\ \hline
\end{tabular}

\begin{tabular}{c|cccccccccccccc}  \hline
\multicolumn{15}{c}{$Y= 0$~~  $I=1          $~~  $J=\frac{1}{2}$~~
$Y_4=\frac{1}{3} $}  \\  \hline
 $D_1D_2J_4I_4$ & 12122 & 12222 & 22122 & 22222 & 24222 & 32122 & 32222 & 34222 & 12224 & 12324 & 22224 & 22324 & 24224 & 24324 \\  \hline
 1~ 7~ 2~ 2 & $ \frac{  1}{   3}$ & $-\frac{  1}{   6}$ & $ 0$ & $ 0$ & $ 0$ & $ \frac{  1}{   6}$ & $-\frac{  1}{  12}$ & $-\frac{  1}{   4}$ & $ 0$ & $ 0$ & $ 0$ & $ 0$ & $ 0$ & $ 0$  \\
 1~ 7~ 2~ 4 & $ 0$ & $ 0$ & $ 0$ & $ 0$ & $ 0$ & $ 0$ & $ 0$ & $ 0$ & $ 0$ & $ 0$ & $ \frac{  1}{   4}$ & $ \frac{  1}{   8}$ & $-\frac{  1}{   4}$ & $-\frac{  3}{   8}$  \\
 1~10~ 3~ 2 & $ 0$ & $ 0$ & $ 0$ & $ 0$ & $ 0$ & $ 0$ & $ 0$ & $ 0$ & $ 0$ & $ 0$ & $ 0$ & $ 0$ & $ 0$ & $ 0$  \\
 1~10~ 3~ 4 & $ 0$ & $ 0$ & $ 0$ & $ 0$ & $ 0$ & $ 0$ & $ 0$ & $ 0$ & $ 0$ & $ 0$ & $ 0$ & $ 0$ & $ 0$ & $ 0$  \\
 2~ 6~ 2~ 2 & $ \frac{  1}{   3}$ & $ \frac{  1}{   6}$ & $ 0$ & $ 0$ & $ 0$ & $ \frac{  1}{   6}$ & $ \frac{  1}{  12}$ & $ \frac{  1}{   4}$ & $ 0$ & $ 0$ & $ 0$ & $ 0$ & $ 0$ & $ 0$  \\
 2~ 6~ 2~ 4 & $ 0$ & $ 0$ & $ 0$ & $ 0$ & $ 0$ & $ 0$ & $ 0$ & $ 0$ & $ 0$ & $ 0$ & $ \frac{  1}{   4}$ & $-\frac{  1}{   8}$ & $-\frac{  1}{   4}$ & $ \frac{  3}{   8}$  \\
\hline
\end{tabular}

\begin{tabular}{c|cccccccccccccc}  \hline
 $D_1D_2J_4I_4$ & 32224 & 32324 & 34224 & 34324 & 12232 & 22232 & 24232 & 32232 & 34232 & 12234 & 12434 & 22234 & 22434 & 24234 \\  \hline
 1~ 7~ 2~ 2 & $ 0$ & $ 0$ & $ 0$ & $ 0$ & $ 0$ & $ 0$ & $ 0$ & $ 0$ & $ 0$ & $ 0$ & $ 0$ & $ 0$ & $ 0$ & $ 0$  \\
 1~ 7~ 2~ 4 & $ 0$ & $ 0$ & $ 0$ & $ 0$ & $ 0$ & $ 0$ & $ 0$ & $ 0$ & $ 0$ & $ 0$ & $ 0$ & $ 0$ & $ 0$ & $ 0$  \\
 1~10~ 3~ 2 & $ 0$ & $ 0$ & $ 0$ & $ 0$ & $ \frac{  1}{   2}$ & $ 0$ & $ 0$ & $ \frac{  1}{   4}$ & $-\frac{  1}{   4}$ & $ 0$ & $ 0$ & $ 0$ & $ 0$ & $ 0$  \\
 1~10~ 3~ 4 & $ 0$ & $ 0$ & $ 0$ & $ 0$ & $ 0$ & $ 0$ & $ 0$ & $ 0$ & $ 0$ & $ 0$ & $ 0$ & $ \frac{  3}{  16}$ & $-\frac{  3}{  16}$ & $ \frac{  1}{  16}$  \\
 2~ 6~ 2~ 2 & $ 0$ & $ 0$ & $ 0$ & $ 0$ & $ 0$ & $ 0$ & $ 0$ & $ 0$ & $ 0$ & $ 0$ & $ 0$ & $ 0$ & $ 0$ & $ 0$  \\
 2~ 6~ 2~ 4 & $ 0$ & $ 0$ & $ 0$ & $ 0$ & $ 0$ & $ 0$ & $ 0$ & $ 0$ & $ 0$ & $ 0$ & $ 0$ & $ 0$ & $ 0$ & $ 0$  \\
 2~ 9~ 3~ 4 & $ 0$ & $ 0$ & $ 0$ & $ 0$ & $ 0$ & $ 0$ & $ 0$ & $ 0$ & $ 0$ & $ \frac{  3}{   8}$ & $ \frac{  1}{  24}$ & $ 0$ & $ 0$ & $ 0$  \\
\hline
\end{tabular}

\begin{tabular}{c|cccccccccccccc}  \hline
 $D_1D_2J_4I_4$ & 24434 & 32234 & 32434 & 34234 & 34434 & 12424 & 22424 & 24424 & 32424 & 34424 & 21222 & 23222 & 21224 & 23224 \\  \hline
 1~ 7~ 2~ 2 & $ 0$ & $ 0$ & $ 0$ & $ 0$ & $ 0$ & $ 0$ & $ 0$ & $ 0$ & $ 0$ & $ 0$ & $ 0$ & $ 0$ & $ 0$ & $ 0$  \\
 1~ 7~ 2~ 4 & $ 0$ & $ 0$ & $ 0$ & $ 0$ & $ 0$ & $ 0$ & $ 0$ & $ 0$ & $ 0$ & $ 0$ & $ 0$ & $ 0$ & $ 0$ & $ 0$  \\
 1~10~ 3~ 2 & $ 0$ & $ 0$ & $ 0$ & $ 0$ & $ 0$ & $ 0$ & $ 0$ & $ 0$ & $ 0$ & $ 0$ & $ 0$ & $ 0$ & $ 0$ & $ 0$  \\
 1~10~ 3~ 4 & $ \frac{  9}{  16}$ & $ 0$ & $ 0$ & $ 0$ & $ 0$ & $ 0$ & $ 0$ & $ 0$ & $ 0$ & $ 0$ & $ 0$ & $ 0$ & $ 0$ & $ 0$  \\
 2~ 6~ 2~ 2 & $ 0$ & $ 0$ & $ 0$ & $ 0$ & $ 0$ & $ 0$ & $ 0$ & $ 0$ & $ 0$ & $ 0$ & $ 0$ & $ 0$ & $ 0$ & $ 0$  \\
 2~ 6~ 2~ 4 & $ 0$ & $ 0$ & $ 0$ & $ 0$ & $ 0$ & $ 0$ & $ 0$ & $ 0$ & $ 0$ & $ 0$ & $ 0$ & $ 0$ & $ 0$ & $ 0$  \\
 2~ 9~ 3~ 4 & $ 0$ & $ \frac{  3}{  16}$ & $ \frac{  1}{  48}$ & $-\frac{  3}{  16}$ & $ \frac{  3}{  16}$ & $ 0$ & $ 0$ & $ 0$ & $ 0$ & $ 0$ & $ 0$ & $ 0$ & $ 0$ & $ 0$  \\
\hline
\end{tabular}

\begin{tabular}{c|cccccccccccccc}  \hline
 $D_1D_2J_4I_4$ & 23132 & 21334 & 23334 & 23424 \\  \hline
 1~ 7~ 2~ 2 & $ 0$ & $ 0$ & $ 0$ & $ 0$  \\
 1~ 7~ 2~ 4 & $ 0$ & $ 0$ & $ 0$ & $ 0$  \\
 1~10~ 3~ 2 & $ 0$ & $ 0$ & $ 0$ & $ 0$  \\
 1~10~ 3~ 4 & $ 0$ & $ 0$ & $ 0$ & $ 0$  \\
 2~ 6~ 2~ 2 & $ 0$ & $ 0$ & $ 0$ & $ 0$  \\
 2~ 6~ 2~ 4 & $ 0$ & $ 0$ & $ 0$ & $ 0$  \\
 2~ 9~ 3~ 4 & $ 0$ & $ 0$ & $ 0$ & $ 0$  \\
\hline
\end{tabular}

\begin{tabular}{c|cccccccccccccc}  \hline
\multicolumn{15}{c}{$Y= 0$~~  $I=1          $~~  $J=\frac{1}{2}$~~
$Y_4=-\frac{2}{3}$}  \\  \hline
 $D_1D_2J_4I_4$ & 12123 & 12223 & 12323 & 22123 & 22223 & 22323 & 24223 & 24323 & 32123 & 32223 & 32323 & 34223 & 34323 & 12233 \\  \hline
 1~ 8~ 2~ 3 & $ \frac{  1}{   9}$ & $-\frac{  1}{   6}$ & $-\frac{  1}{  18}$ & $ 0$ & $ 0$ & $ 0$ & $ 0$ & $ 0$ & $ \frac{  1}{  18}$ & $-\frac{  1}{  12}$ & $-\frac{  1}{  36}$ & $-\frac{  1}{   4}$ & $-\frac{  1}{   4}$ & $ 0$  \\
 2~ 7~ 2~ 3 & $ \frac{  4}{   9}$ & $ 0$ & $ \frac{  1}{  18}$ & $ 0$ & $ 0$ & $ 0$ & $ 0$ & $ 0$ & $ \frac{  2}{   9}$ & $ 0$ & $ \frac{  1}{  36}$ & $ 0$ & $ \frac{  1}{   4}$ & $ 0$  \\
 2~10~ 3~ 3 & $ 0$ & $ 0$ & $ 0$ & $ 0$ & $ 0$ & $ 0$ & $ 0$ & $ 0$ & $ 0$ & $ 0$ & $ 0$ & $ 0$ & $ 0$ & $ \frac{  1}{   2}$  \\
 3~ 6~ 2~ 3 & $ \frac{  1}{   9}$ & $ \frac{  1}{   6}$ & $-\frac{  1}{  18}$ & $ 0$ & $ 0$ & $ 0$ & $ 0$ & $ 0$ & $ \frac{  1}{  18}$ & $ \frac{  1}{  12}$ & $-\frac{  1}{  36}$ & $ \frac{  1}{   4}$ & $-\frac{  1}{   4}$ & $ 0$  \\
 5~ 7~ 2~ 3 & $ 0$ & $ 0$ & $ 0$ & $ 0$ & $ \frac{  1}{   2}$ & $ 0$ & $ \frac{  1}{   2}$ & $ 0$ & $ 0$ & $ 0$ & $ 0$ & $ 0$ & $ 0$ & $ 0$  \\
 5~ 7~ 3~ 3 & $ 0$ & $ 0$ & $ 0$ & $ 0$ & $ 0$ & $ 0$ & $ 0$ & $ 0$ & $ 0$ & $ 0$ & $ 0$ & $ 0$ & $ 0$ & $-\frac{  1}{   6}$  \\
\hline
\end{tabular}

\begin{tabular}{c|cccccccccccccc}  \hline
 $D_1D_2J_4I_4$ & 22233 & 24233 & 32233 & 34233 & 21223 & 23223 & 21333 & 23133 & 23333 \\  \hline
 1~ 8~ 2~ 3 & $ 0$ & $ 0$ & $ 0$ & $ 0$ & $ 0$ & $ 0$ & $ 0$ & $ 0$ & $ 0$  \\
 2~ 7~ 2~ 3 & $ 0$ & $ 0$ & $ 0$ & $ 0$ & $ 0$ & $ 0$ & $ 0$ & $ 0$ & $ 0$  \\
 2~10~ 3~ 3 & $ 0$ & $ 0$ & $ \frac{  1}{   4}$ & $-\frac{  1}{   4}$ & $ 0$ & $ 0$ & $ 0$ & $ 0$ & $ 0$  \\
 3~ 6~ 2~ 3 & $ 0$ & $ 0$ & $ 0$ & $ 0$ & $ 0$ & $ 0$ & $ 0$ & $ 0$ & $ 0$  \\
 5~ 7~ 2~ 3 & $ 0$ & $ 0$ & $ 0$ & $ 0$ & $ 0$ & $ 0$ & $ 0$ & $ 0$ & $ 0$  \\
 5~ 7~ 3~ 3 & $ 0$ & $ 0$ & $-\frac{  1}{  12}$ & $-\frac{  3}{   4}$ & $ 0$ & $ 0$ & $ 0$ & $ 0$ & $ 0$  \\
 5~10~ 2~ 3 & $ 0$ & $ 0$ & $ 0$ & $ 0$ & $ 0$ & $ 0$ & $ 0$ & $ 0$ & $ 0$  \\
\hline
\end{tabular}

\begin{tabular}{c|cccccccccccccc}  \hline
\multicolumn{ 7}{c}{$Y= 0$~~  $I=0          $~~  $J=\frac{5}{2}$~~
$Y_4=\frac{1}{3} $}  \\  \hline
 $D_1D_2J_4I_4$ & 12214 & 22214 & 24414 & 32214 & 34414 & 23314 \\  \hline
 4~ 7~ 1~ 4 & $ \frac{  1}{   2}$ & $ 0$ & $ 0$ & $ \frac{  1}{   4}$ & $-\frac{  1}{   4}$ & $ 0$  \\
 5~ 6~ 1~ 4 & $ 0$ & $-\frac{  1}{   4}$ & $ \frac{  3}{   4}$ & $ 0$ & $ 0$ & $ 0$  \\
 6~ 5~ 1~ 4 & $ \frac{  1}{   6}$ & $ 0$ & $ 0$ & $ \frac{  1}{  12}$ & $ \frac{  3}{   4}$ & $ 0$  \\
 6~ 7~ 1~ 4 & $-\frac{  1}{   3}$ & $ 0$ & $ 0$ & $ \frac{  2}{   3}$ & $ 0$ & $ 0$  \\
 7~ 4~ 1~ 4 & $ 0$ & $-\frac{  3}{   4}$ & $-\frac{  1}{   4}$ & $ 0$ & $ 0$ & $ 0$  \\
 7~ 6~ 1~ 4 & $ 0$ & $ 0$ & $ 0$ & $ 0$ & $ 0$ & $-1$  \\ \hline
\end{tabular}

\begin{tabular}{c|cccccccccccccc}  \hline
\multicolumn{ 6}{c}{$Y= 0$~~  $I=0          $~~  $J=\frac{5}{2}$~~
$Y_4=-\frac{2}{3}$}  \\  \hline
 $D_1D_2J_4I_4$ & 12215 & 22215 & 24415 & 32215 & 34415 \\  \hline
 4~ 8~ 1~ 5 & $ \frac{  1}{   3}$ & $ 0$ & $ 0$ & $ \frac{  1}{   6}$ & $-\frac{  1}{   2}$  \\
 5~ 7~ 1~ 5 & $ 0$ & $-\frac{  1}{   2}$ & $ \frac{  1}{   2}$ & $ 0$ & $ 0$  \\
 7~ 5~ 1~ 5 & $ \frac{  1}{   3}$ & $ 0$ & $ 0$ & $ \frac{  1}{   6}$ & $ \frac{  1}{   2}$  \\
 7~ 7~ 1~ 5 & $-\frac{  1}{   3}$ & $ 0$ & $ 0$ & $ \frac{  2}{   3}$ & $ 0$  \\
 8~ 4~ 1~ 5 & $ 0$ & $-\frac{  1}{   2}$ & $-\frac{  1}{   2}$ & $ 0$ & $ 0$  \\ \hline
\end{tabular}

\begin{tabular}{c|cccccccccccccc}  \hline
\multicolumn{15}{c}{$Y= 0$~~  $I=0          $~~  $J=\frac{3}{2}$~~
$Y_4=\frac{1}{3} $}  \\  \hline
 $D_1D_2J_4I_4$ & 12224 & 12324 & 22224 & 22324 & 24224 & 24324 & 32224 & 32324 & 34224 & 34324 & 12214 & 22214 & 24414 & 32214 \\  \hline
 1~ 7~ 2~ 4 & $ 0$ & $ 0$ & $ \frac{  1}{   4}$ & $ \frac{  1}{   8}$ & $-\frac{  1}{   4}$ & $-\frac{  3}{   8}$ & $ 0$ & $ 0$ & $ 0$ & $ 0$ & $ 0$ & $ 0$ & $ 0$ & $ 0$  \\
 2~ 6~ 2~ 4 & $ 0$ & $ 0$ & $ \frac{  1}{   4}$ & $-\frac{  1}{   8}$ & $-\frac{  1}{   4}$ & $ \frac{  3}{   8}$ & $ 0$ & $ 0$ & $ 0$ & $ 0$ & $ 0$ & $ 0$ & $ 0$ & $ 0$  \\
 4~ 7~ 1~ 4 & $ 0$ & $ 0$ & $ 0$ & $ 0$ & $ 0$ & $ 0$ & $ 0$ & $ 0$ & $ 0$ & $ 0$ & $ \frac{  1}{   2}$ & $ 0$ & $ 0$ & $ \frac{  1}{   4}$  \\
 4~ 7~ 2~ 4 & $ 0$ & $ 0$ & $ \frac{  3}{   8}$ & $ 0$ & $ \frac{  3}{   8}$ & $ 0$ & $ 0$ & $ 0$ & $ 0$ & $ 0$ & $ 0$ & $ 0$ & $ 0$ & $ 0$  \\
 4~10~ 2~ 4 & $ 0$ & $ \frac{  1}{   4}$ & $ 0$ & $ 0$ & $ 0$ & $ 0$ & $ 0$ & $ \frac{  1}{   8}$ & $ 0$ & $-\frac{  1}{   8}$ & $ 0$ & $ 0$ & $ 0$ & $ 0$  \\
 5~ 6~ 1~ 4 & $ 0$ & $ 0$ & $ 0$ & $ 0$ & $ 0$ & $ 0$ & $ 0$ & $ 0$ & $ 0$ & $ 0$ & $ 0$ & $-\frac{  1}{   4}$ & $ \frac{  3}{   4}$ & $ 0$  \\
\hline
\end{tabular}

\begin{tabular}{c|cccccccccccccc}  \hline
 $D_1D_2J_4I_4$ & 34414 & 12424 & 22424 & 24424 & 32424 & 34424 & 23314 & 21224 & 23224 & 23424 \\  \hline
 1~ 7~ 2~ 4 & $ 0$ & $ 0$ & $ 0$ & $ 0$ & $ 0$ & $ 0$ & $ 0$ & $ 0$ & $ 0$ & $ 0$  \\
 2~ 6~ 2~ 4 & $ 0$ & $ 0$ & $ 0$ & $ 0$ & $ 0$ & $ 0$ & $ 0$ & $ 0$ & $ 0$ & $ 0$  \\
 4~ 7~ 1~ 4 & $-\frac{  1}{   4}$ & $ 0$ & $ 0$ & $ 0$ & $ 0$ & $ 0$ & $ 0$ & $ 0$ & $ 0$ & $ 0$  \\
 4~ 7~ 2~ 4 & $ 0$ & $ 0$ & $ \frac{  1}{   8}$ & $ \frac{  1}{   8}$ & $ 0$ & $ 0$ & $ 0$ & $ 0$ & $ 0$ & $ 0$  \\
 4~10~ 2~ 4 & $ 0$ & $-\frac{  1}{   6}$ & $ 0$ & $ 0$ & $-\frac{  1}{  12}$ & $-\frac{  1}{   4}$ & $ 0$ & $ 0$ & $ 0$ & $ 0$  \\
 5~ 6~ 1~ 4 & $ 0$ & $ 0$ & $ 0$ & $ 0$ & $ 0$ & $ 0$ & $ 0$ & $ 0$ & $ 0$ & $ 0$  \\
 5~ 6~ 2~ 4 & $ 0$ & $ \frac{  1}{   4}$ & $ 0$ & $ 0$ & $ \frac{  1}{   8}$ & $-\frac{  3}{   8}$ & $ 0$ & $ 0$ & $ 0$ & $ 0$  \\
\hline
\end{tabular}

\begin{tabular}{c|cccccccccccccc}  \hline
\multicolumn{15}{c}{$Y= 0$~~  $I=0          $~~  $J=\frac{3}{2}$~~
$Y_4=-\frac{2}{3}$}  \\  \hline
 $D_1D_2J_4I_4$ & 12225 & 22225 & 24225 & 32225 & 34225 & 12215 & 22215 & 24415 & 32215 & 34415 & 12425 & 22425 & 24425 & 32425 \\  \hline
 2~ 7~ 2~ 5 & $ 0$ & $ \frac{  1}{   2}$ & $-\frac{  1}{   2}$ & $ 0$ & $ 0$ & $ 0$ & $ 0$ & $ 0$ & $ 0$ & $ 0$ & $ 0$ & $ 0$ & $ 0$ & $ 0$  \\
 4~ 8~ 1~ 5 & $ 0$ & $ 0$ & $ 0$ & $ 0$ & $ 0$ & $ \frac{  1}{   3}$ & $ 0$ & $ 0$ & $ \frac{  1}{   6}$ & $-\frac{  1}{   2}$ & $ 0$ & $ 0$ & $ 0$ & $ 0$  \\
 4~ 8~ 2~ 5 & $ 0$ & $ \frac{  1}{   4}$ & $ \frac{  1}{   4}$ & $ 0$ & $ 0$ & $ 0$ & $ 0$ & $ 0$ & $ 0$ & $ 0$ & $ 0$ & $ \frac{  1}{   4}$ & $ \frac{  1}{   4}$ & $ 0$  \\
 5~ 7~ 1~ 5 & $ 0$ & $ 0$ & $ 0$ & $ 0$ & $ 0$ & $ 0$ & $-\frac{  1}{   2}$ & $ \frac{  1}{   2}$ & $ 0$ & $ 0$ & $ 0$ & $ 0$ & $ 0$ & $ 0$  \\
 5~ 7~ 2~ 5 & $-\frac{  1}{   6}$ & $ 0$ & $ 0$ & $-\frac{  1}{  12}$ & $ \frac{  1}{   4}$ & $ 0$ & $ 0$ & $ 0$ & $ 0$ & $ 0$ & $ \frac{  1}{   6}$ & $ 0$ & $ 0$ & $ \frac{  1}{  12}$  \\
 5~10~ 2~ 5 & $ 0$ & $ 0$ & $ 0$ & $ 0$ & $ 0$ & $ 0$ & $ 0$ & $ 0$ & $ 0$ & $ 0$ & $ 0$ & $ \frac{  1}{   2}$ & $-\frac{  1}{   2}$ & $ 0$  \\
\hline
\end{tabular}

\begin{tabular}{c|cccccccccccccc}  \hline
 $D_1D_2J_4I_4$ & 34425 & 21225 & 23225 & 23425 \\  \hline
 2~ 7~ 2~ 5 & $ 0$ & $ 0$ & $ 0$ & $ 0$  \\
 4~ 8~ 1~ 5 & $ 0$ & $ 0$ & $ 0$ & $ 0$  \\
 4~ 8~ 2~ 5 & $ 0$ & $ 0$ & $ 0$ & $ 0$  \\
 5~ 7~ 1~ 5 & $ 0$ & $ 0$ & $ 0$ & $ 0$  \\
 5~ 7~ 2~ 5 & $-\frac{  1}{   4}$ & $ 0$ & $ 0$ & $ 0$  \\
 5~10~ 2~ 5 & $ 0$ & $ 0$ & $ 0$ & $ 0$  \\
 7~ 2~ 2~ 5 & $ 0$ & $ 0$ & $ 0$ & $ 0$  \\
 7~ 5~ 1~ 5 & $ 0$ & $ 0$ & $ 0$ & $ 0$  \\
 7~ 5~ 2~ 5 & $ 0$ & $ 0$ & $ 0$ & $ 0$  \\
 7~ 7~ 1~ 5 & $ 0$ & $ 0$ & $ 0$ & $ 0$  \\
 7~ 7~ 2~ 5 & $ 0$ & $ \frac{  1}{   3}$ & $-\frac{  2}{   3}$ & $ 0$  \\
 7~10~ 2~ 5 & $ 0$ & $ 0$ & $ 0$ & $ 0$  \\
 8~ 4~ 1~ 5 & $ 0$ & $ 0$ & $ 0$ & $ 0$  \\
 8~ 4~ 2~ 5 & $ \frac{  1}{   4}$ & $ 0$ & $ 0$ & $ 0$  \\
 8~ 9~ 2~ 5 & $ 0$ & $ \frac{  1}{   3}$ & $ \frac{  1}{   6}$ & $-\frac{  1}{   2}$  \\
 9~ 8~ 2~ 5 & $ 0$ & $ 0$ & $ 0$ & $ 0$  \\
10~ 5~ 2~ 5 & $-\frac{  1}{   2}$ & $ 0$ & $ 0$ & $ 0$  \\
10~ 7~ 2~ 5 & $ 0$ & $-\frac{  1}{   3}$ & $-\frac{  1}{   6}$ &
$-\frac{  1}{   2}$  \\ \hline
\end{tabular}%

\end{tiny}

\begin{center}
Table A2. The index of diquark cluster

%
\end{center}

\end{document}